\documentstyle[prb,aps,multicol,epsf]{revtex}
\renewcommand{\narrowtext}{\begin{multicols}{2} \global\columnwidth20.5pc}
\renewcommand{\widetext}{\end{multicols} \global\columnwidth42.5pc}
\begin{document}
\draft
\title{
An overview on the phase diagram of the frustrated two-leg ladder model\\
}
\author{T. Hakobyan\cite{tigran}, J.H.~Hetherington\cite{jack} 
and M.~Roger\cite{michel}}
\address{
 Service de Physique de l'Etat Condens\'e,
Commissariat \`a l'Energie Atomique,\\ Centre d'Etudes de Saclay,
91191 Gif sur Yvette Cedex, France.\\
}
\date{\today}
\maketitle

\begin{abstract}
Using Density-Matrix Renormalization Group, we investigate the general 
phase diagram of the frustrated two-leg ladder with Heisenberg interactions 
along legs, rungs and diagonals. We confirm that all antiferromagnetic
gapped states belong to the same universality class as the Haldane phase.
In a three-dimensional phase-diagram, we determine a continuous surface 
with singularities in the string-order parameter or  its first derivative,
corresponding to a transition between two Haldane phases with different
topological order.
Some parts of this transition surface are critical  with zero gap
and vanishing string-order parameter. In the complementary parts, 
the transition is first order with finite gap and string order.
The boundary of this surface with the ferromagnetic region is 
a critical end line, when the surface is critical and a triple line
anywhere else. Part of this boundary coincides with the exactly soluble 
model proposed by D. V. Dmitriev, V. Ya Krivnov and A. A. Ovchinnikov 
[Phys. Rev. B {\bf 56}, 5985 (1997)].

\end{abstract}

\narrowtext

\section{Introduction}

${\cal N}$-legged spin ladders are formed by assembling ${\cal N}$ 
spin-$1\over 2$ 
chains one next to the other. Many weakly-coupled ladder systems have
now been synthetized. Among them, the family $Sr_{n-1}Cu_{n+1}O_{2n}$ consists
in weakly coupled ${1\over 2}(n+1)$-legged ladders which are obtained from
the $CuO_2$ planes of the parent compound $SrCuO_2$.
After the suggestion that mechanisms similar to those occurring in the $CuO_2$
planes of cuprate ceramics may also lead to superconductivity\cite{dagotto},
an intense theoretical and experimental activity has been developed on these
systems. The doped spin-ladder compound $Sr_{14-x}Ca_{x}Cu_{24}O_{41}$ is 
superconducting up to 10~K under pressure.
Some theoretical tools which are specific to one-dimensional systems make the 
theoretical understanding of ladders easier than that of the $CuO_2$ planes 
responsible for the high-temperature superconductivity in cuprate ceramics. 
It is believed that the mechanisms leading to superconductivity are similar 
in both systems and the understanding of spin ladders should give some 
insight in the physics of the more complex two-dimensional superconducting 
systems. 

A first step in the theoretical understanding of these systems is the study of
undoped ladders in which the spin interactions are of Heisenberg type. The 
exact solution of the spin-$1\over 2$ Heisenberg chain has been known for 
seventy years from the Bethe ansatz\cite{bethe}. 
The quantum fluctuations prevent 
long-range  antiferromagnetic order, there is no gap in the excitation
spectrum and spin-spin correlations decay in power law as a function of the 
distance. However, the crossover from ladders to the square lattice, by 
increasing the number of coupled chains, is far from being smooth. While
ladders with an odd number of legs exhibit properties
similar to those of a single chain, i.e. gapless excitations and a power-law
fall-off of spin-spin correlations, ladders with an even number of legs have a
finite energy gap to the lowest spin-1 excitation and exponential decay of
spin-spin correlations. These remarkable quantum properties are remininiscent
of those, first conjectured by Haldane\cite{haldane}, 
concerning the spin-$n\over 2$
Heisenberg chain with, respectively, n odd and n even. Therefore, a 
natural question arises: how are the ladder phases related to the phases in 
the spin chain? 
\begin{figure}[htbp]
\epsfxsize=7truecm
\epsfbox{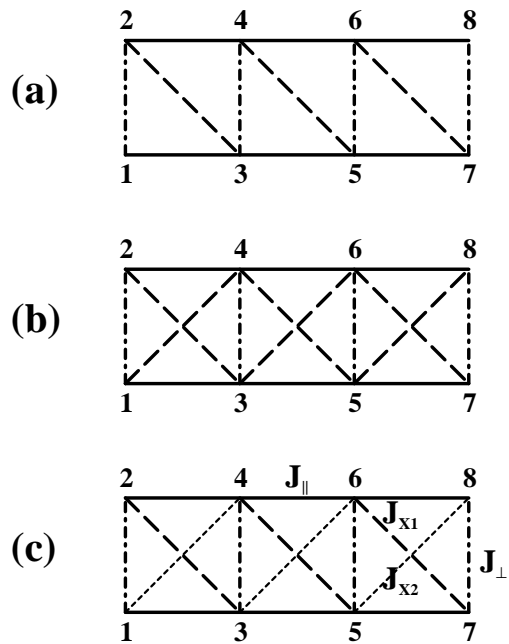}
\caption{Frustrated-ladder models: (a) The ``Zig-Zag chain''. (b) Frustrated
ladder with equal diagonal interactions. (c) The more general model studied
in this paper}
\end{figure}
For the two-leg ladder, this question has been partly answered by 
White\cite{white1}. He considered a two-leg ladder with additional interaction 
along one diagonal of each plaquette [Fig.~1~(a)] and showed by varying this 
interaction that there is a continuous path from the Haldane to the ladder 
phase which does not cross any phase boundary or critical points. Some 
controversial arguments were published recently by Wang\cite{wang}. He 
considered a ladder with additional equal interactions along both diagonals of 
each plaquette [Fig.~1~(b)]. Using Density-Matrix Renormalization Group (DMRG) 
calculations\cite{white} he proved the existence of a first-order transition 
line and suggested the occurrence of a phase-transition from the Haldane-phase
to a gapped singlet phase with different sensitivity with respect to boundary
conditions.

It is therefore interesting to study the general ``frustrated two-leg
ladder model'' [Fig.~1~(c)], including two different diagonal interactions
in the plaquettes, which contains both previous works as particular
cases and to draw a general phase diagram.
We thus consider the following Hamiltonian:

\begin{eqnarray}
H=2\sum_{i=1}^{L} [ &J_\Vert ({\bf S_{i}\cdot S_{i+2}} +
            {\bf S_{2i-1}\cdot S_{2i+1}}) +	 \nonumber\\	
           &J_\perp {\bf S_{2i-1}\cdot S_{2i}} + \hfill \nonumber\\
	   &J_{X1} {\bf S_{2i}\cdot S_{2i+1}} +
	   J_{X2} {\bf S_{2i-1}\cdot S_{2i+2}}]
\end{eqnarray}
($L$ is the ladder length and the number of spins is $N=2L$)
and propose  a general three-dimensional phase diagram for this model.
There is an obvious symmetry of this Hamiltonian by exchanging $J_{X1}$ and 
$J_{X2}$ and it is more convenient to use:
\begin{eqnarray}
   {\cal S}=J_{X1}+J_{X2}& \qquad {\rm and} \nonumber \\
   {\cal D}=J_{X1}-J_{X2}&
\end{eqnarray}
as parameters. We choose the following energy scaling:
\begin{equation}
 |2J_\Vert | + |J_\perp| + |{\cal S}| + |{\cal D}| = 1
\end{equation}
Due to the symmetry, it is sufficient to consider ${\cal D}>0$. We choose as 
independent variables $x=2J_\Vert$, $y={\cal S}$, $z=J_\perp$ and ${\cal D}$ 
is given by the previous relation. The variables $\{x,y,z\}$ obey:
$|x|+|y|+|z| < 1$ and are thus contained in the regular octahedron 
represented in Fig.~2. 

Using Density-Matrix Renormalization Group (DMRG), Conformal Field Theory and
some exact analytical results, we draw transition surfaces separating various
phases and identify their critical parts.
We review the most important results obtained earlier for particular cases, 
including the spin-1 and spin-$1\over 2$ Heisenberg chain, the frustrated
spin-1 and spin-$1\over 2$ chains with interactions between nearest and
next nearest neighbors, the dimerized chain, and the usual ladder, which are
all gathered in this general phase diagram. We do not put any restriction
on the sign of the four exchange parameters and provide a new insight in
the regions with negative (ferromagnetic) interactions. 
\begin{figure}[htbp]
\epsfxsize=7truecm
\epsfbox{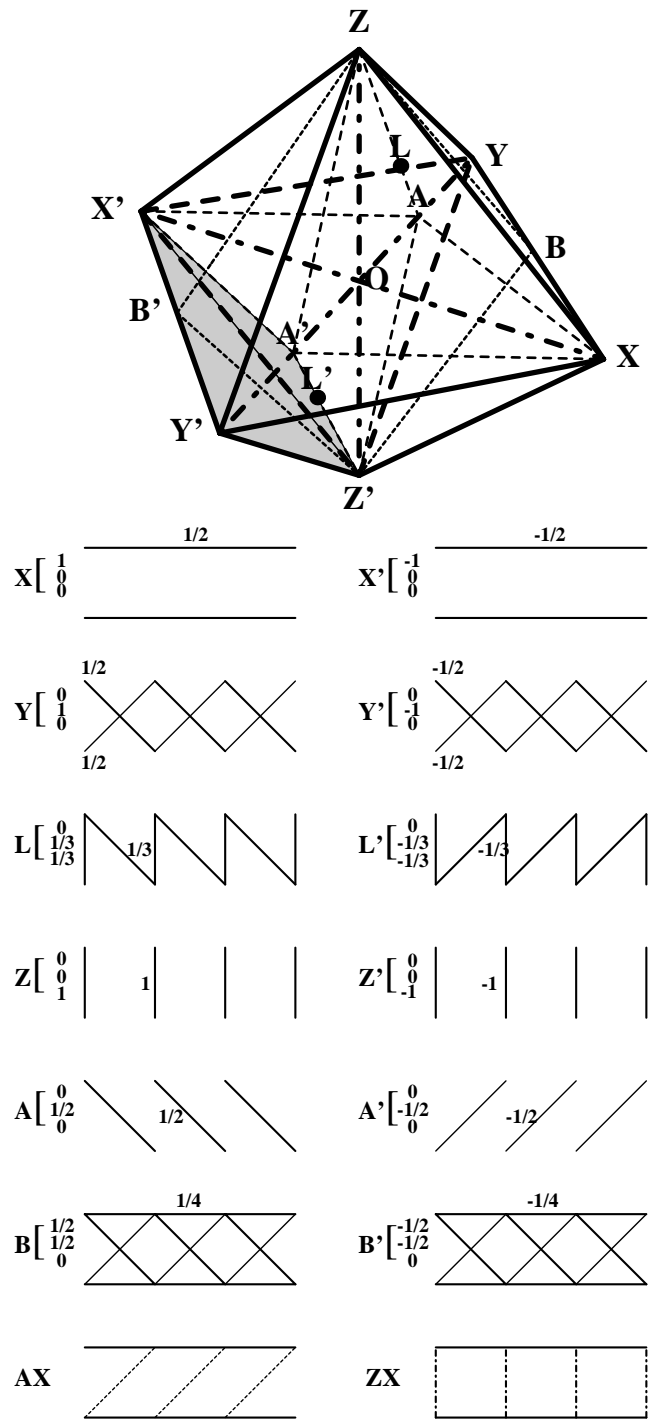}
\caption{The regular octahedron inscribed in a sphere of radius 1, 
representing the parameter-space of our Hamiltonian [Eq. (1)] with the energy
scaling  of Eq. (3). Some remarkable points with their 
 $\{x,y,z\}$ coordinates are listed: the spin-$1\over 2$
Heisenberg antiferromagnetic (X,Y,L) or ferromagnetic (X',Y',L') chain,
products of independent dimers (A,Z) or triplet pairs (A',Z'), the
usual ladder (AX and ZX segments) etc... The ``Zig-Zag'' chain
represented in Fig.~1~(a)
corresponds to the surface of the irregular octahedron (AA',XX',ZZ'). In the
gray volume, all interactions are negative (ferromagnetic).}
\end{figure}
We begin with the study of the vicinity of particular points
of this octahedron where the low-lying eigenstates can be mapped onto
those of the Haldane spin-1 or spin-$1\over 2$ chain (section II), 
and introduce three different phases: two topologically distinct gapped states
and the ferromagnetic phase. In our three-dimensional phase diagram,
the points X, Y, L representing the  spin-$1\over 2$ antiferromagnetic
Heisenberg chain appear as particular critical points lying in zero-gap
critical surfaces separating the two gapped Haldane phases.
In section III, we briefly describe the method that we use to determine
transition surfaces between different phases: a Density-Matrix 
Renormalization group algorithm, with ``suitable'' open boundary conditions.

Section IV is devoted to particular planes in the octahedron corresponding 
the the ladder model represented in Fig. 1 (a). This model can also be viewed
as a ``Zig-Zag'' Heisenberg chain with nearest and next-nearest neighbor
interactions and has already been extensively studied.
We briefly review the most important results obtained 
for this particular case.
Using   Lancz\"os diagonalisation and DMRG,
we  investigate in section V the whole phase diagram in the octahedron and 
draw transition surfaces between the three different phases 
identified in section II. We show that no other phase appears.

\section{Identification of different phases of the phase-diagram}

\subsection{Evidence for two topologically distinct spin-1 Haldane phases}

The points Z' (all interactions null, except $J_\perp =-1$) and 
A' (all interactions null except $J_{X2}=-1/2$) represent independent pairs
of spins 1/2 coupled ferromagnetically along rungs and along
plaquette diagonals respectively. The $3^{L}$ degenerate ground-state is
identical to that of $L$ independent spins-1: 
${\bf \tilde S_i}=({\bf S_{2i}} + {\bf S_{2i-1}})$ at Z' and
${\bf \tilde S_i}=({\bf S_{2i}} + {\bf S_{2i-3}})$ at A'. In the neighborhood
of both points,  degenerate perturbation theory is relevant as far as
the other coupling parameters are much smaller than the gap between the
$3^L$ degenerate ground-state and the first excited state (i.e. $<<1$). The
split $3^L$ lowest levels can be mapped onto the states of a chain of
L interacting spins-1 with effective Hamiltonian (at first order in the
coupling parameters):
\begin{equation}
H_{eff}={2J_\Vert+J_{X1}+J_{X2}\over 2}
\sum_{i=1}^{L}{\bf \tilde S_i \cdot \tilde S_{i+1}}
\end{equation}
in the neighborhood of point Z' and
\begin{equation}
H_{eff}={2J_\Vert+J_\perp\over 2}
\sum_{i=1}^{L}{\bf \tilde S_i \cdot \tilde S_{i+1}}+
{{J_{X1}}\over{2}}\sum_{i=1}^{L}{\bf \tilde S_i \cdot \tilde S_{i+2}}
\end{equation}
in the neighborhood of point A'

The Hamiltonian (4) for $(2J_\Vert+J_{X1}+J_{X2})>0$ and the
 Hamiltonian (5) for
$2J_\Vert > 0$ and $J_{X1}=0$ ($J_{X1}=0$ is satisfied in the planes A'XZ, 
A'XZ', A'X'Z) represent the spin-1 Haldane chain. For $J_{X1}=0$ both
spin-1 Hamiltonians (4) and (5) are formally identical. However, 
they represent
different topological configurations of the original ladder:  the spin-1
operators ${\bf \tilde S_i}$ represent the projection on the triplet state of 
the sum of a pair of spins-$1\over 2$ along one rung in Hamiltonian (4) 
and along one diagonal of a plaquette in Hamiltonian (5).

The most important breakthrough for the understanding of the antiferromagnetic
spin-1 Heisenberg chain has been the discovery by 
Affleck-Kennedy-Lieb-Tasaki (AKLT)\cite{aklt} of an 
exactly soluble Hamiltonian
differing from the Heisenberg model only by addition to the Hamiltonian of
a biquadratic term: 
\begin{equation}
H_{AKLT}=\sum_i\left[ {\bf \tilde S_i \cdot \tilde S_{i+1}}
+{1\over 3}({\bf \tilde S_i\cdot \tilde S_{i+1}})^2\right]
\end{equation}
Its exact ground-state called ``Valence-Bond Solid'' (VBS) is constructed out 
of valence bonds\cite{anderson}, it is nondegenerate and breaks no symmetry.
It can be represented as schematized in Fig. 3 (a). 
\begin{itemize}
\item
The spin-1 variables are expressed as a sum of two spin-$1\over 2$ variables
\item
Let $\psi_{\alpha,\beta}$ denote a spin-1 state in terms of the symmetrized
spin-$1\over 2$ variables:
\begin{eqnarray}
\psi_{++}=&|++>, \nonumber\\
\psi_{+-}=&\psi_{-+}=(|+-> + |-+>)/\sqrt{2} \\
\psi_{--}=&|-->. \nonumber
\end{eqnarray}
\item
Each spin-$1\over 2$ is contracted with one spin-$1\over 2$
at a neighbor site into a singlet state: a ``valence bond''. 
Contraction of two spin-$1\over 2$ variables into a singlet
state can be written as $\epsilon^{\alpha\beta}|\alpha\beta>/\sqrt{2}$, where
$\epsilon^{\alpha,\beta}$ represents the antisymmetric tensor: $\epsilon^{++}=
\epsilon^{--}=0$ and $\epsilon^{+-}=-\epsilon^{-+}=1$
\item
The contractions are such that every spin-1 is connected to both its neighbors
by a valence bond and the VBS state can be written:
\begin{eqnarray}
|\Psi_{VBS}>=2^{-N/2}\psi_{\alpha_1\beta_1}\epsilon^{\beta_1\alpha_2}
\psi_{\alpha_2\beta_2}\epsilon^{\beta_2\alpha_3}\cdots \nonumber\\
\psi_{\alpha_i\beta_i}\epsilon^{\beta_i\alpha_{i+1}}\cdots
\psi_{\alpha_L\beta_L}\epsilon^{\beta_L\alpha_1}
\end{eqnarray}
\end{itemize}
 
The VBS state, used as variational state for the Heisenberg spin-1 chain, gives
an upper bound of the energy which is only a few percent higher than the
best approximates through DMRG and is thought to capture the essentials of
the physics of the this model.
The VBS state has some ``hidden'' topological long-range 
order\cite{rommelse}. 
\begin{figure}[htbp]
\epsfxsize=7truecm
\epsfbox{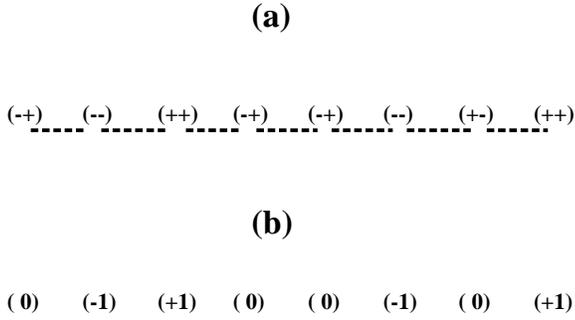}
\caption{Schematic view of a typical component of the VBS state described by
Eq. 8. (a) The dashed links represent 
Valence-Bonds between two fictive neighboring
spins-$1\over 2$. (b) Sequence of spin-1 $S^z$ values corresponding to (a):
the non-zero spins are in a N\'eel order (-1,1,-1,1,....)} 
\end{figure}
It is easy to see that each term of the sum in the 
previous relation has $S_i^z$
sequences ordered in a way shown schematically in Fig.~3~(b). 
If we discard the 
sites with $S_i^z =0$, the remaining $S_i^z =\pm 1$ sites have 
N\'eel order: $|1,-1,1,-1,1,-1,\cdots>$.
This topological order is characterized by the non-local ``string-order
parameter'':
\begin{equation}
{\cal O}^z (|m-l|) = \left< \tilde S_l^z 
\exp \left( i \pi \sum_{k=l+1}^{k=m-1} \tilde S_k^z\right) \tilde S_m^z\right> 
\end{equation} 
A normalized string-order parameter:
\begin{equation}
\tilde{\cal O}^z (d)={-{\cal O}^z (d)\over < (\tilde S^z)^2 >^2}
\end{equation}
has been later introduced by White\cite{white}. Its value for $d\to\infty$
is 1 for the VBS state and 0.84 for the Heisenberg spin-1 chain.

From the previous discussion about the neighborhoods of points Z' and A'
in the phase diagram, it is natural to introduce two string-order parameters:
\begin{itemize}
\item
{${\cal O}_\perp (d)$ obtained with $\tilde S_i^z = (S_{2i}^z+S_{2i-1}^z)$
in Eq.~9, with triplet pairs of spins along rungs }
\item
{${\cal O}_X (d)$ obtained with $\tilde S_i^z = (S_{2i}^z+S_{2i-3}^z)$
in Eq.~9, with triplet pairs of spins along  one diagonal of each plaquette}
\end{itemize}
These two topologically distinct Haldane phases, that we now call 
$HD_\perp$ and $HD_X$, characterized respectively by
the string-order parameters ${\cal O}_\perp$ and ${\cal O}_X$ have
been introduced in a recent work by Kim et al.\cite{kim}, with slightly
different notations. They are schematized in Fig.~4.
Using field-theoretical methods, Kim et al. have found
evidences for some transition lines between these two Haldane phases around
some particular points of the phase diagram corresponding to two-decoupled
Heisenberg chains (see sections IV and V).

With $J_{X1}$ strictly positive, the full Hamiltonian (5) represents the
``frustrated'' antiferromagnetic Heisenberg S=1 chain, with interactions
$J_1=(2J_\Vert + J_\perp )/4 $ and $J_2=J_{X1}/4$ between first and second
neighbors. This model has been studied by Kolezhuk et al.\cite{kolezhuk}
They find a sharp discontinuity in the string-order parameter ${\cal O}_X$
at $J_2/J_1 \approx 0.7444$, suggesting a first-order transition. 
The significance of this first-order transition will appear in section V,
when placed in the framework of the general phase-diagram.

\subsection{The antiferromagnetic spin-1/2 Heisenberg chain}

In Fig. 2a, the points $X$: ($2J_\Vert=1$, all other $J$'s null)
and $Y$: ($J_{X1}=J_{X2}=1/2$, all other $J$'s null) represent two decoupled
spin-$1\over 2$  antiferromagnetic Heisenberg chains, and the point 
$L$ ($J_\perp=J_{X1}=1/3$, all other $J$'s null)
represents a single spin-$1\over 2$
antiferromagnetic Heisenberg chain. The exact 
solution of this Hamiltonian is known from the Bethe Ansatz\cite{bethe}.
There is no gap in the excitation spectrum and the spin-spin correlations
decrease as power law in terms of the distance. In  sections IV and V, we
shall prove that these points lie in critical surfaces separating the
two Haldane phases with different topological order.  

\subsection{The ferromagnetic state}

In the grey volume A'Y'X'Z' of the octahedron [Fig.~2~(a)], all interactions
are negative. The ground-state is ferromagnetic. It has gapless 
spin-wave excitations. The ferromagnetic state obviously extends beyond
this volume. Its limits will be determined in sections IV and V.
\begin{figure}[htbp]
\epsfxsize=7truecm
\epsfbox{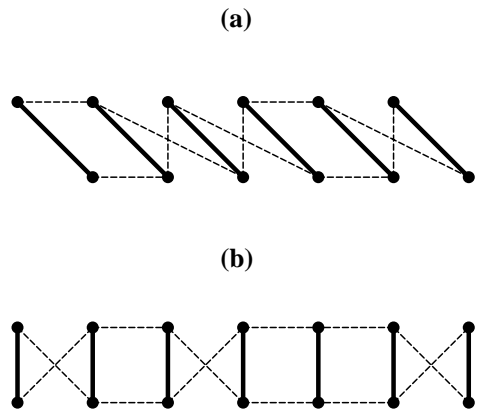}
\caption{The two topologically distinct Haldane phases $HD_X$ (a)
and $HD_\perp$ (b). The heavy links represent triplet pairing and
the dashed links schematize Valence Bonds.}
\end{figure}

\begin{figure}[htbp]
\epsfxsize=7truecm
\epsfbox{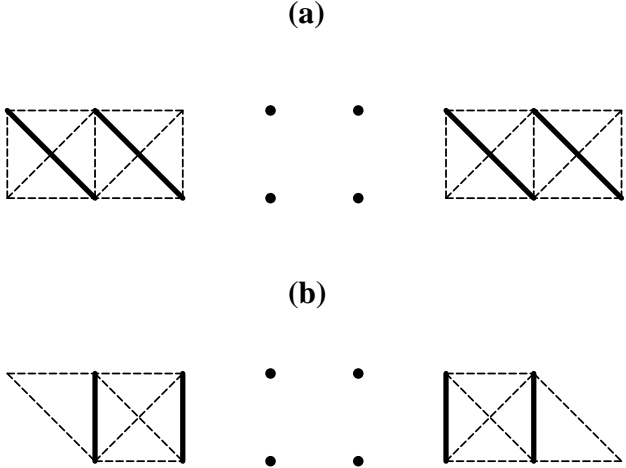}
\caption{Growing algorithms used in DMRG calculation with free boundaries:
(a) in the $HD_X$ Haldane Phase, where triplet pairing occurs along
diagonals, (b) in the $HD_\perp$ Haldane phase, where triplet pairing 
occurs along rungs. In either case, the ground state is non degenerate.}
\end{figure}
\section{DMRG with suitable boundary conditions}

Our purpose is to determine transition lines or surfaces separating different
phases, in particular the two topologically distinct Haldane phases 
$HD_\perp$ and $HD_X$ defined in section II. We use the
Density-Matrix Renormalization Group (DMRG) method with open boundary 
conditions which gives much more accurate results than with periodic
boundary conditions\cite{white}. However, a well-known inconvenience for
the Haldane spin-1 chain with open boundary conditions is the four-fold
degeneracy of the ground-state. This degeneracy is easily understood
through the description in Fig.~3~(a) and Eq.~(8) of the VBS state, which is
thought to capture the essential of the physics of the Haldane chain. 
In Fig.~3~(a), with open boundary conditions, there are two spin-$1\over 2$
at each
end of the chain which are not contracted through a valence bond to any other
half-spin. 
The interaction between these extra half-spins fall off 
exponentially with L and in the long-chain limit, there is a four-fold
degeneracy of the ground-state due to these two free spins.

In their DMRG treatment of the spin-1 chain, White and Huse remove this 
inconvenient degeneracy by adding at each end an extra spin-$1\over 2$ 
which forms
a singlet with half of the last spin-1 in the chain\cite{spin1}. When
applying DMRG to the frustrated ladder problem, the adaptation of this
trick is straightforward and already built in the ladder topology.
 In the $HD_X$ phase, we use the boundary conditions shown in Fig.~5~(a) which
are generally chosen in treating the ladder problem\cite{wang}. In the
$HD_\perp$ phase, we use other boundary conditions schematized in
Fig.~5~(b). If the ground state is viewed as a VBS state, in either case
Figs.~5 (a) and (b) show that there is no ``free spin'' (``free'' means
 not linked to any
another spin with a ``valence bond'') at either end. Consequently, the
ground-state is non degenerate in the long chain limit. In either case
our DMRG algorithm for growing the chain length proceeds by adding one
rung to each of the right and left Blocks as shown in Fig. 5. 
What differs is the shape of the initial Block: it is a rectangle in case (a)
for the $HD_X$ phase and a trapezo\"\i d in case (b) for the $HD_\perp$
phase.

 Using the same algorithm for any point of the phase-diagram (most 
previous DMRG studies on ladders have proceeded in such a way\cite{wang}) 
generally leads to numerical difficulties when crossing the phase-boundary. 
Suppose that we start from a point in the $HD_X$ phase with the suitable
algorithm [Fig.~5~(a)] and follow a path to the $HD_\perp$ phase, at the
transition to the $HD_\perp$ phase, the ground-state, which was
non-degenerate becomes four-fold degenerate and we need to keep a much
larger number of states to obtain a reasonable approximation of the transition
point. To determine numerically many transition points in
a three-dimensional phase diagram, we need a ``cheap'' algorithm for
calculating a transition point with a reasonable accuracy in a minimum of 
computer time.

Our algorithm uses the fact that at a phase transition, there are some
singularities in the first or second derivatives of the energy. There is 
a discontinuity in the first derivative if the transition is first order.
Here, if the transition is second order the second derivative is generally
infinite at the critical point. We thus proceed as follows:
\begin{itemize}
\item
We choose a path $P_0P'_0$  going from the $HD_X$ phase $(P_0)$ to the 
$HD_\perp$ phase $(P'_0)$ and parametrized by the variable u. Using the 
algorithm schematized in Fig.~5~(a), we calculate at $P_0$ the energy and its 
first and second derivatives. The first derivative of the energy, with respect
to $x=2J_\Vert$, $y={\cal S}$ and $z=J_\perp$ is obtained by calculating the 
correlations:
\begin{eqnarray}
C_\Vert =& <1+4 {\bf S_{i}\cdot S_{i+2}}>/2 \nonumber\\
C_\perp=& < 1+4{\bf S_{2i-1}\cdot S_{2i}}>/2 \nonumber\\
C_{X1}=& <1+4{\bf S_{2i}\cdot S_{2i+1}}>/2 \\
C_{X2}=& < 1+4{\bf S_{2i-1}\cdot S_{2i+2}}>/2 \nonumber
\end{eqnarray}
The derivative of the energy $\partial E_0/\partial u$ along the path
is obtained from the preceding correlations. The second derivative
 $\partial^2 E_0/\partial u^2$ is obtained numerically through the same
calculation with an infinitesimal variation on u (chosen at the square root
of the numerical precision). An approximate second order polynomial expression
of the energy near $P_0$ is deduced:
\begin{equation}
E= E_0 + s  \partial E_0/\partial s + (s^2/2) \partial^2 E_0/\partial s^2
\end{equation}
\item
Using the other algorithm schematized in Fig.~5~(b), a polynomial approximation
of the energy around $P'_0$ is obtained in the same way. 
\item
A first approximation $ P^C_0$ of the critical point is given by the 
intersection of both polynomials.
\item
A new starting point $P_1$ is chosen half way between $P_0$ and $P^C_0$,
a new end point $P'_1$ is chosen half way between $P^C_0$ and $P'_0$
and the process is reiterated.
\end{itemize} 
An accuracy of 4 to 5 digits is generally obtained for the critical point
after 8 to 10 iterations. We keep 30 to 60 states and use chain 
length $L\approx 50$. Some  accuracy tests we be given in section V.

\section{The ``Zig-Zag chain''}

The ladder model, with only one diagonal interaction, represented in 
Fig.~1~(a) can also be viewed as a spin-$1\over 2$ 
dimerized next-nearest neighbor Heisenberg chain. Using the scaling from
Eq. (3), the two-dimensional parameter space for this model corresponds to
the eight faces of the irregular octahedron XX'AA'ZZ' (Fig. 6). There is
an obvious symmetry, in this model, obtained by exchanging rung and
diagonal interaction. The symmetry line XLX'L'X, corresponding to equal rung 
and diagonal interactions splits the parameter surface into two pieces with
a one to one correspondence. Due to our non-trivial energy scaling [Eq. (3)],
this mapping does not correspond however to simple reflections.

The AXZ face, in which all interactions are antiferromagnetic 
has been studied for many years.
On the symmetry line XL, the point G corresponding to $J_\perp=J_{X1}=2J_\Vert$
has been considered by Majumdar and Ghosh\cite{mg} thirty years ago. The
ground state is doubly degenerate and corresponds to a product of singlet
pairs (or dimers) along  rungs or diagonals. Shastry and 
Sutherland\cite{shastry} have later proven that the dimer state along the
rungs (resp. diagonals) is the exact ground state on the 
whole line GZ (resp. GA). 
Note that dimer states have perfect ``string order'', characterized by the
normalized order parameters\cite{white} $\tilde{\cal O}_{X} =1$ on GZ and  
$\tilde{\cal O}_\perp =1$ on GA, respectively. Following the path AGZ, the
string order parameter $\tilde{\cal O}_\perp$ jumps abruptly from 1 to 0 at
the Majumdar-Ghosh point G and $\tilde{\cal O}_\perp$ jumps respectively
from 0 to one. The whole symmetry line LX is a transition 
line\cite{chitra,neugebauer1,emery}. According to our notations introduced in
section II, it corresponds to a phase transition between the two symmetrical 
Haldane phases $HD_X$ and $HD_\perp$.
Both ends X and L correspond to the spin-$1\over 2$ Heisenberg chain. 
The vicinities of these points have been studied
through field-theoretical methods\cite{kim,emery,eggert}. The transition is 
first order near X and second order near L.
On XL the nature of the transition changes from first to second order at
a tricritical point T. From field-theoretical results, at T the
spectrum develops  a gap between the S=0 ground state and the first excited
S=1 states and simultaneously the degeneracy between the two lowest
S=0 and S=1 excited states is split. Using Lancz\"os diagonalisation on
finite systems, Emery et al\cite{emery}
 have extrapolated in the thermodynamic limit
the exchange parameter values at which the energy difference between the
lowest S=0 and S=1 excited states vanishes and found T at 
 $J_\Vert /J_\perp \approx 0.241$. Using the same method, we have found
no evidence for another tricritical point on the symmetry line LX' in the
AX'Z face. Hence TLX' is a critical line  probably up to  X'.

In the  A'X'Z' face, all interactions are negative and the ground 
state is ferromagnetic. The ferromagnetic phase extends further in the 
neighboring faces: A'X'Z, A'XZ' and AX'Z'. In these faces, the line of 
transition from ferromagnetic to antiferromagnetic state corresponds to the 
exactly solvable Hamiltonian studied by Dmitriev, Krivnov and 
Ovchinnikov\cite{ovchinnikov} and hereafter referred as the DKO
model Hamiltonian:
\begin{eqnarray}
 H_{DKO}=&-\sum_{i=1}^L \left({\bf S_{2i-1}\cdot S_{2i}} -{1\over 4}\right)
  \nonumber\\
&-(\nu -1)\sum_{i=1}^L \left({\bf S_{2i}\cdot S_{2i+1}} -{1\over 4}\right)  
\nonumber\\
&+{(\nu -1)\over 2\nu}\sum_{i=1}^L\left({\bf S_{i}\cdot S_{i+2}}-{1\over 4}
\right)
\end{eqnarray}

The ferromagnetic ground-state of this Hamiltonian is degenerate with a
singlet state, which can be written explicitly\cite{ovchinnikov}. With
the scaling defined in Eq. (3), we can write the line of transition
in the following parametric form:
\begin{itemize}
\item{A'X'Z face}
\begin{eqnarray}
x=& 2J_\vert = (\nu-1)/(1+2\nu- \nu^2) \nonumber\\
y=& {\cal S}= J_{X2}= -\nu /(1+2\nu- \nu^2) \nonumber\\
z=& J_\perp = -\nu (\nu -1)/(1+2\nu- \nu^2)
\end{eqnarray}
with $0<\nu <1$
\item{AX'Z' face}
\begin{eqnarray}
x=& 2J_\vert =(\nu-1)/(1+2\nu -\nu^2) \nonumber\\
y=& {\cal S}= J_{X1}=-\nu (\nu-1)/(1+2\nu -\nu^2) \nonumber\\
z=& J_\perp = -\nu /(1+2\nu -\nu^2)
\end{eqnarray}
with $0<\nu <1$
\item{A'XZ' face}

The line XL' corresponding to $J_{X2}=J_\perp$ is a symmetry line. On the
upper side of this line the ferromagnetic transition is parametrized by
\begin{eqnarray}
x=& 2J_\Vert = (\nu -1)/(\nu^2 + 2\nu -1) \nonumber\\
y=& {\cal S}= J_{X2} = -\nu /(\nu^2 + 2\nu -1) \nonumber\\
z=& J_\perp = -\nu (\nu -1)/(\nu^2 + 2\nu -1)
\end{eqnarray}
with $1<\nu <2$. On the other side, we have:

\begin{figure}[htbp]
\epsfxsize=7truecm
\epsfbox{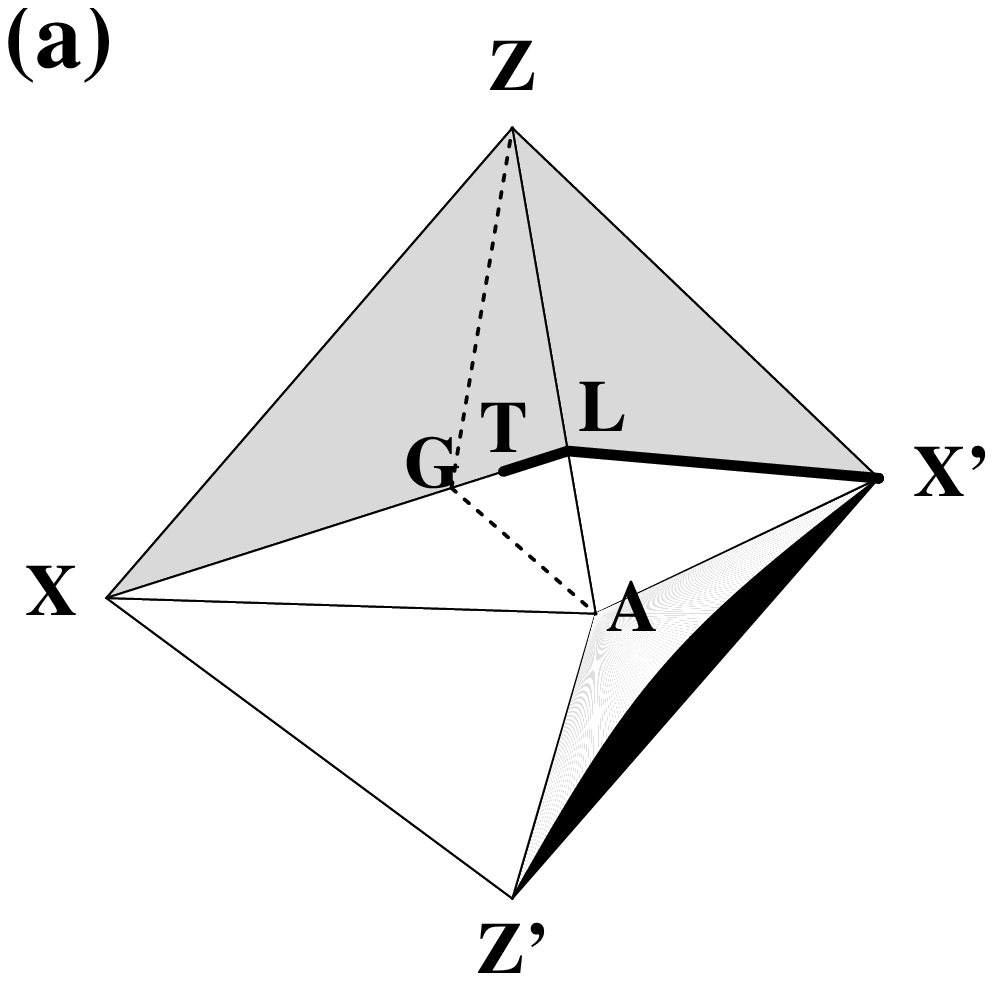}
\end{figure}
\begin{figure}[htbp]
\epsfxsize=7truecm
\epsfbox{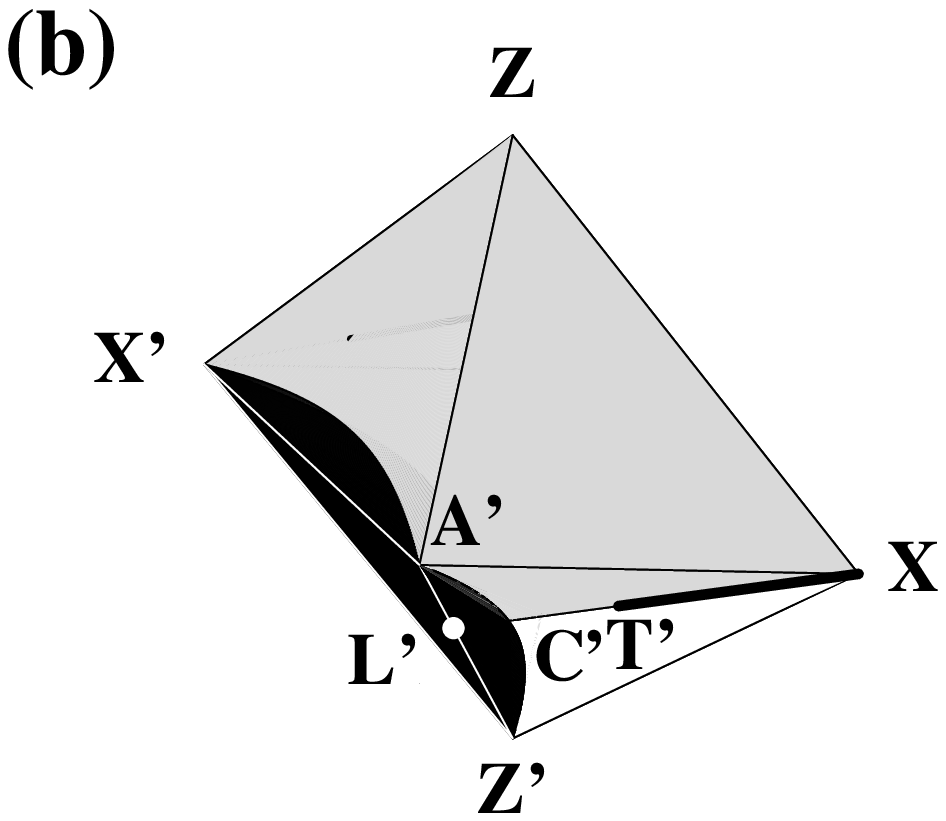}
\caption{
The surface of the irregular octahedron XX'AA'ZZ' represents the 
two-dimensional phase diagram of the ``Zig-Zag chain'' model [Fig.~1~(a)]. 
The symmetry line XLX'L'X corresponding to equal rung and diagonal 
interactions splits this surface into two parts which map each other through
the exchange of rung and diagonal. Above (resp. below) this line, the 
antiferromagnetic gray (resp. white) area corresponds to the  $HD_X$ 
(resp.  $HD_\perp$) Haldane phase. The black area represents the ferromagnetic
state. TLX' and T'X are two zero-gapped critical lines corresponding to
second order transition between the two Haldane phases. On TX and T'E', the
transition is first order. G represents the
Majumdar-Ghosh point. From Shastry and Sutherland, on the thin dotted
segments GA and GZ, the ground-state is exact and corresponds to a product
of independent dimers}
\end{figure}
\begin{eqnarray}
x=& 2J_\Vert = (\nu-1)/(2\nu^2 -1)  \nonumber\\
y=& {\cal S}= J_{X2}= -\nu (\nu -1)/(2\nu^2 -1)  \nonumber\\
z=& J_\perp = -\nu /(2\nu^2 -1)
\end{eqnarray}
with $1<\nu <2$
\end{itemize}

In the A'XZ' phase, with ferromagnetic rung and diagonal interactions, the 
symmetry line XL' meets the ferromagnetic region at C' (1/7,-2/7,-2/7).
From field-theoretical results\cite{kim}, near X this line is critical
with zero gap and corresponds to a second order phase transition between
the two symmetric Haldane phases $HD_\perp$ and $HD_X$. Using Lancz\"os
diagonalizations on finite systems with N up to 24 particles and the
method previously described\cite{emery} we have found that this critical
line ends at a tricritical point T' corresponding to $2J_\Vert\approx 0.42$.
From T' to C', the transition between the two symmetrical gapped phases
is first order and C' appears as a triple point.

Figure 6 summarizes the results:
\begin{itemize}
\item
The X'LTXT'C' part of the symmetry line (equal rung and diagonal interactions)
corresponds to a phase transition between two symmetric Haldane phases
$HD_\perp$ and $HD_X$ corresponding through the exchange of rung and diagonal.
\item
X'LT and XT' (heavy segments in Fig. 6) are critical lines with zero gap
and second order transition while XT and C'T' corresponds to a first order
transition with degenerate ground state and finite gap.
\item
On either side of the critical line, the $HD_\perp$ and $HD_X$ phase
correspond through a simple symmetry. Consequently, the topology
of their ground-states differ, but they have the same thermodynamic properties
and same critical exponents near the critical line. 
In the usual sense of ``universality'' refering to critical exponents, 
they belong to the same universality class\cite{note}.
\end{itemize}

Our study of the general ladder model [Fig.~1~(c)] will now extend this
simple diagram in three dimensions. Critical lines will be generalized into
critical surface etc...

\section{The general phase-diagram}

\subsection{Some exact critical lines and surfaces}

In the preceding section, we have seen that in the surface of the 
irregular octahedron
XX'AA'ZZ' corresponding to the ``Zig-Zag'' chain the ferromagnetic
boundary is exactly known. We shall extend here this exact boundary to
a part of the whole ferromagnetic volume. We have put forward the lines
GZ and GA
where a dimer state is the exact ground state. We shall prove that the exact
dimer state extends over some planar surface and find exactly the boundary
between this surface and the ferromagnetic phase.

Let us rewrite the Hamiltonian [Eq.~(1)] as a sum over plaquettes: 
\begin{eqnarray}
H=&\sum_{i=1}^L H^\Box _i  \qquad {\rm with} \nonumber\\
H^\Box _i =
&2J_\Vert ({\bf S_{i}\cdot S_{i+2}} +
            {\bf S_{2i-1}\cdot S_{2i+1}}) +	 \nonumber\\	
           &J_\perp {\bf S_{2i-1}\cdot S_{2i}} + \hfill \nonumber\\
	   &2J_{X1} {\bf S_{2i}\cdot S_{2i+1}} +
	   2J_{X2} {\bf S_{2i-1}\cdot S_{2i+2}}]
\end{eqnarray}
The eigenvalues of $H^\Box _i$ for an isolated plaquette are:
\begin{eqnarray}
\lambda_1=&-3J_\perp /2 \nonumber\\
\lambda_2=& J_\perp /2 + J_{X1}+J_{X2}  \nonumber\\
\lambda_3=& J_\perp /2 -2( J_{X1}+J_{X2}) \\
\lambda_4=& -J_\perp /2 \nonumber\\
\lambda_5=& -{1\over 2}(J_{X1}+J_{X2}+[J_\perp ^2 +5J_{X1}^2\nonumber\\
          &-6J_{X1}J_{X2}+5J_{X2}^2
             -2J_\perp (J_{X1}+J_{X2})]^{1\over 2} ) \nonumber\\
\lambda_6=& -{1\over 2}(J_{X1}+J_{X2}-[J_\perp ^2 +5J_{X1}^2 \nonumber\\
          &-6J_{X1}J_{X2}+5J_{X2}^2
             -2J_\perp (J_{X1}+J_{X2})]^{1\over 2} ) \nonumber
\end{eqnarray}
The eigenvalue $\lambda_1$ corresponds to a product of rung dimers. It is
straightforward to check that it is an exact eigenstate of the total 
Hamiltonian $H$ if $2J_\Vert = {\cal S}=J_{X1}+J_{X2}$. 
The eigenvalue $\lambda_2$ 
corresponds to the ferromagnetic state, it is always an exact eigenstate of
the full Hamiltonian H. A lower bound to the ground-state  energy per unit
length $E_0/L$ of the full Hamiltonian $H$ is given by the
lowest eigenvalue $\lambda_{min}$ of $H^\Box _i$. If the lowest eigenvalue
is $\lambda_2$, the ground state is ferromagnetic. The volume where
$\lambda_{min}=\lambda_2$ provides an inner envelope to the ferromagnetic 
region. After  little algebra, we find that this inner envelope is defined 
by the surfaces:
\begin{equation}
w=4 J_{X1} J_{X2} + (J_{X1}+J_{X2})(2J_\Vert +J_\perp) + 2 J_\Vert J_\perp =0
\end{equation}
inside the irregular tetrahedron XX'AA'ZZ' and
\begin{equation}
w'=(J_{X1}+J_{X2})(2J_\Vert +J_\perp) + 2 J_\Vert J_\perp =0
\end{equation}
outside this tetrahedron.

On another hand, an outer envelope of the ferromagnetic region is provided
by a simple calculation of spin-waves instabilities in the ferromagnetic
phase. For the ``Zig-Zag chain'', Kolezhuk and Mikeska\cite{km} have already 
pointed out that a simple spin-wave instability calculation in the 
ferromagnetic region gives exactly the DKO transition line.
Introducing two kinds of bosons for representing the spin operators on two
different legs, we find one optical and one acoustic mode. The energy of the
acoustic mode at $k\to 0$ is:
\begin{equation}
E_{sw}^{ac}=-wk^2
\end{equation}
where $w$ corresponds to the expression given in Eq.~(20)
The ferromagnetic state is instable with respect to spin-waves
 for $w>0$. The surface $w=0$ thus provides an outer
envelope to the ferromagnetic region. Inside the irregular octahedron
XX'AA'ZZ', we have thus provided an outer and an inner envelope which 
coincide. The surface $w=0$ [Eq. (20)] is thus an exact boundary of the 
ferromagnetic region.

It is also worth paying some particular attention to the plane ZBZ'B' of the
regular octahedron (see Fig. 2) 
which corresponds to $2J_\Vert =(J_{X1}+J_{X2})$.
As pointed out above, in that plane, the dimer-state with independent dimers 
along the rungs is an exact eigenstate of the Hamiltonian.

As previously
emphasised this dimer state has perfect string order $\tilde{\cal O}_X=1$ and
corresponds to the $HD_X$ Haldane phase. Its exact first order transition
line with the ferromagnetic phase is given by the equation:
\begin{equation}
\lambda_1=\lambda_2 \qquad {\it i.e.} \qquad 2J_\perp+J_{X1}+J_{X2}=0
\end{equation}
from point D=(-2/5,-2/5,1/5) to point $C=(-2/7,-2/7,1/7)$ corresponding
to DKO model (see Fig.~7). The other part CZ' of the
transition line from the ferromagnetic to the $HD_\perp$ phase is given exactly
by Eq.~(20).

The line ZBZ' with $J_{X1}=J_{X2}=J_\Vert >0$ corresponds to the so called 
``composite-spin model''\cite{composite}. The Hamiltonian can be written as:
\begin{eqnarray}
H=& H_1 + H_2 \qquad {\rm with:} \nonumber\\
H_1=& 2J_\Vert \sum_{i=1}^L {\bf \tilde S_i\cdot \tilde S_{i+1}} \nonumber\\
H_2=& 2J_\perp \sum_{i=1}^L {\bf S_{2i-1}\cdot S_{2i}}
\end{eqnarray}
where ${\bf \tilde S_i= S_{2i-1}+S_{2i}}$ represents the sum of a pair of
spins-1/2 on a rung. Hamiltonians $H_1$ and $H_2$ commute. The ground state
$E_2^0$ of $H_2$ corresponds to a product of dimers along rungs
i.e: $E_2^0/L = -3J_\perp /2 $. It is easy to check that it is also an 
eigenstate of $H$ with eigenvalue $E_2^0$.
The states with all pairs  ${\bf \tilde S_i= S_{2i-1}+S_{2i}}$ along
the rungs in a triplet state are degenerate eigenstates of $H_2$ with 
eigenvalues $E_2=LJ_\perp /2$. The subspace spanned by these $3^L$ states
corresponds to the spin-1 chain. The low-energy spectrum of $H_1$ corresponds
to that of the spin-1 Haldane chain, and the eigenstates of $H_1$ are also
eigenstates of $H$. The ground state of the spin-1 Haldane chain has an
energy $E_1^0\approx -1.401484039\times 2J_\Vert$ from DMRG 
calculations\cite{spin1} and is an eigenstate of $H$ with energy 
$E_1=E_1^0 + E_2$. Starting from the dimer phase at point Z, we have a first
order transition to the Haldane phase $ HD_\perp$ at the crossing between 
these two energy levels $E_1^0 + E_2=E_2^0$ corresponding to a critical
value $J_\perp/J_\Vert \approx 1.401484039$ (point $\rm H$). 
The Dimer phase has 
perfect string order $\tilde {\cal O}_X=1$ corresponding to the 
$HD_X$ phase. At the transition, the order parameter 
$\tilde{\cal O}_X$ jumps from 1 to 0, while $\tilde {\cal O}_\perp$ jumps from
 0 to 0.84, the value corresponding to the spin-1 antiferromagnetic chain.

\begin{figure}[htbp]
\epsfxsize=7truecm
\epsfbox{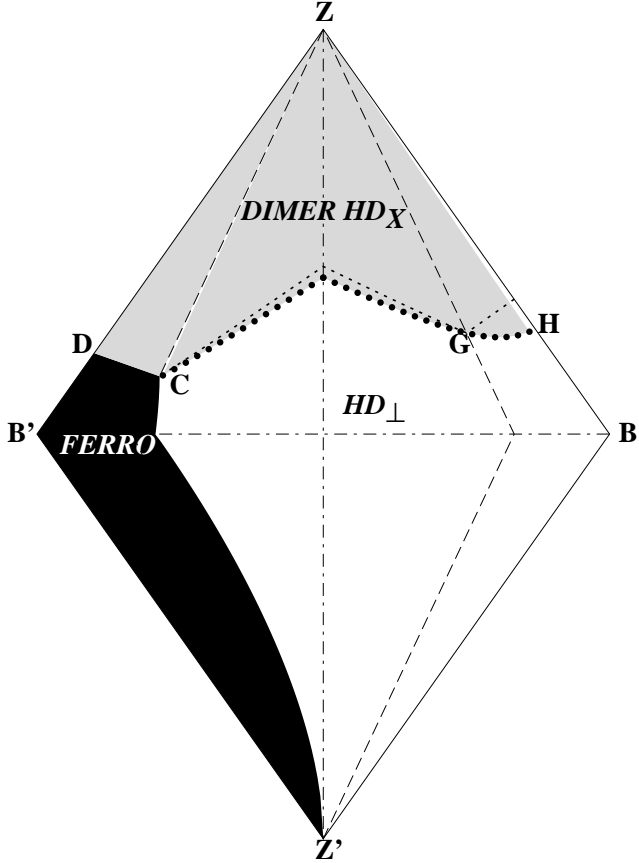}
\caption{The $(J_{X1}+J_{X2})=2J_\Vert$ plane. The dashed lines correspond
the intersection of this plane with the ``Zig-Zag'' model and G represents
the Majumdar-Ghosh point. The black area represents the ferromagnetic
region, whose boundaries are exactly known. The line Z'BZ represents the
``composite-spin'' model:
$J_{X1}=J_{X2}=J_\Vert$. At $H$, there is a first order transition from
the Haldane $HD_\perp$ phase to an exact Dimer phase which has the same 
topological long-range order as the $HD_X$ phase. 
The heavy dotted line CGH represents the first order
transition between the Dimer Haldane phase $HD_X$ (grey area) and the
Haldane $HD_\perp$ phase.
The thin dotted line is an exact inner bound to the exact dimer phase.
$C=(-2/7,-2/7,1/7)$ is an exact Triple point.}
\end{figure}

We have calculated the whole transition line $CGH$ between the
$HD_\perp$ phase and $HD_X$ dimer phase through the DMRG scheme
explained in section III. The transition is first order all along this
line and the point C =(-2/7,-2/7,1/7) appears as an exact triple point.
Note that the Majumdar-Ghosh point G belongs to this line and is also an
exact transition point. From C to G a trivial but remarkable
exact inner bound to the Dimer state is obtained when the dimer energy
$\lambda_1$ ceased to be the lowest eigenvalue for an isolated plaquette,
i.e when $\lambda_1 = \lambda_6$ (see Eq. 20). With our energy
scaling [Eq.~(3)], this lower bound corresponds to
\begin{equation}
J_\perp = (-2+5{\cal S}+ \sqrt{8-36{\cal S}+41{\cal S}^2})/2
\end{equation}
for ${\cal S}>0$ and
\begin{equation}
J_\perp = (-2-3{\cal S}+ \sqrt{8+28{\cal S}+25{\cal S}^2})/2
\end{equation}
for ${\cal S}<0$
(see the thin dotted line in Fig. 7).
Beyond G, The upper bound to the dimer phase corresponds to the intersection
of $\lambda_1$ and $\lambda_3$, i.e. to the segment $J_\perp ={\cal S}$.

\subsection{The general phase diagram}
We shall consider successively the 8 sectors of the tetrahedron XX'YY'ZZ' with 
the notation $[\epsilon_x,\epsilon_y,\epsilon_z]$, where 
$\epsilon_\alpha =\pm$ represents the sign of the coordinate $\alpha$ 
\subsubsection{The $[+,+,+]$ sector, with $J_\Vert$, $J_\perp$ and
               ${\cal S}$ positive}

\smallskip

\noindent {\it a. The boundary face XYZ}

\smallskip
The boundary face XYZ, with equal antiferromagnetic diagonal interactions
 $J_{X1}=J_{X2}=J_X$ [Fig.~1~(b)], has been studied by Wang\cite{wang}. 
There is an obvious symmetry which exchanges $J_\Vert$ and $J_X$ 
by twisting every other rung  by an angle $\pi$ around 
the axis of the ladder. ZB is a symmetry line. It is thus sufficient to study 
the half triangle XBZ. 
From the recent field-theoretical results of Kim et al.\cite{kim}, near X, 
which represents two decoupled spin-$1\over 2$ chains, 
there is a first order transition
between the two Haldane phases $HD_X$ and $HD_\perp$ 
along the line $J_\perp = {\cal S}$. 
From the previous paragraph there is a first
order transition between the same phases along the symmetry axe ZB at $H$.
Using the DMRG algorithm described in section III, we have calculated the
full transition line from X to $H$. It is show in Fig. 8. It is in good
agreement with earlier results obtained by Wang\cite{wang}. However,
in contrast to the conclusions of Wang,  our results
show that all along this line we have a first-order transition
with finite gap from the
$HD_X$  to the $HD_\perp$ Haldane phase, in 
agreement with theoretical-field results\cite{kim}.

 As pointed
out in section III, in contrast to our algorithm which uses different boundary
conditions on either side of the transition line to avoid  the
degeneracy of the ground state, Wang has used everywhere the DMRG
algorithm with boundary conditions corresponding to Fig.~5~(a).
\begin{figure}[htbp]
\epsfxsize=7truecm
\epsfbox{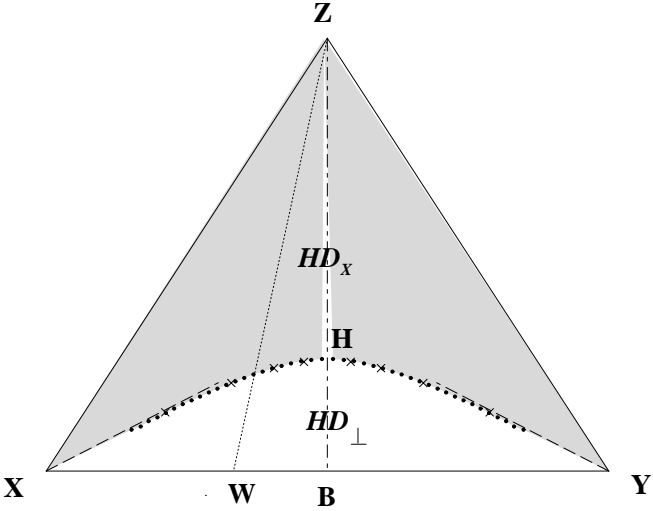}
\caption{The XYZ face with equal diagonal interactions and all $J$'s positive.
The symmetry axis ZB corresponds to the ``composite spin'' model with a 
first-order transition from the 
Haldane $HD_\perp$ phase to the dimer $HD_X$ phase
at $H$. The dashed lines are exact first order transition lines between the
 $HD_\perp$ and  $HD_X$ phase in the neighborhoods of X and Y (two independent
spin-$1\over 2$ chains) obtained through field-theoretical methods). The dots
are our first order transition points obtained through DMRG. They are compared
to earlier results of Wang (crosses). The behavior of nearest-neighbor pair
correlations, gap and string-order parameter along the lines ZB and ZW
corresponding to $J_\Vert= J_{X1}+J_{X2}$ are shown in Fig.~9}
\end{figure}
 As he observed, below the
transition line, the ground state is  degenerate and it is nondegenerate
above this line. However, his conclusion suggesting a transition from a
Haldane phase to a ``Dimer'' phase with different properties concerning the
behavior with respect to boundary conditions is incorrect. 
If we use everywhere the DMRG algorithm with the other boundary condition
corresponding to Fig.~5~(b) 
we find a degenerate ground state above the transition
line and a non-degenerate ground state below! The correct interpretation
is, as previously stressed, the transition between two topologically distinct
Haldane phases. Typical sets of curves showing the discontinuities in the
first-neighbor pair correlations [Eq.~(11)], gap and string-order parameters
along the dotted line ZW in Fig.~8 corresponding to $J_\Vert= J_{X1}+J_{X2}$
and along the symmetry line ZB are shown in Fig.~9.

\smallskip
\noindent {\it b. The OYZ plane}
\smallskip

The OYZ $(J_\Vert =0)$ plane corresponds to coupled dimerized chains. 
 In this plane (Fig.~10) the point L corresponds
to a simple antiferromagnetic Heisenberg chain and Y represents two decoupled
Heisenberg chains. Strong arguments have been given for the existence of a
critical line joining L to Y\cite{sierra1,sierra2} and a simple analytic
approximation of this line has also been conjectured\cite{sierra2}.

\begin{figure}[htbp]
\epsfxsize=7truecm
\epsfbox{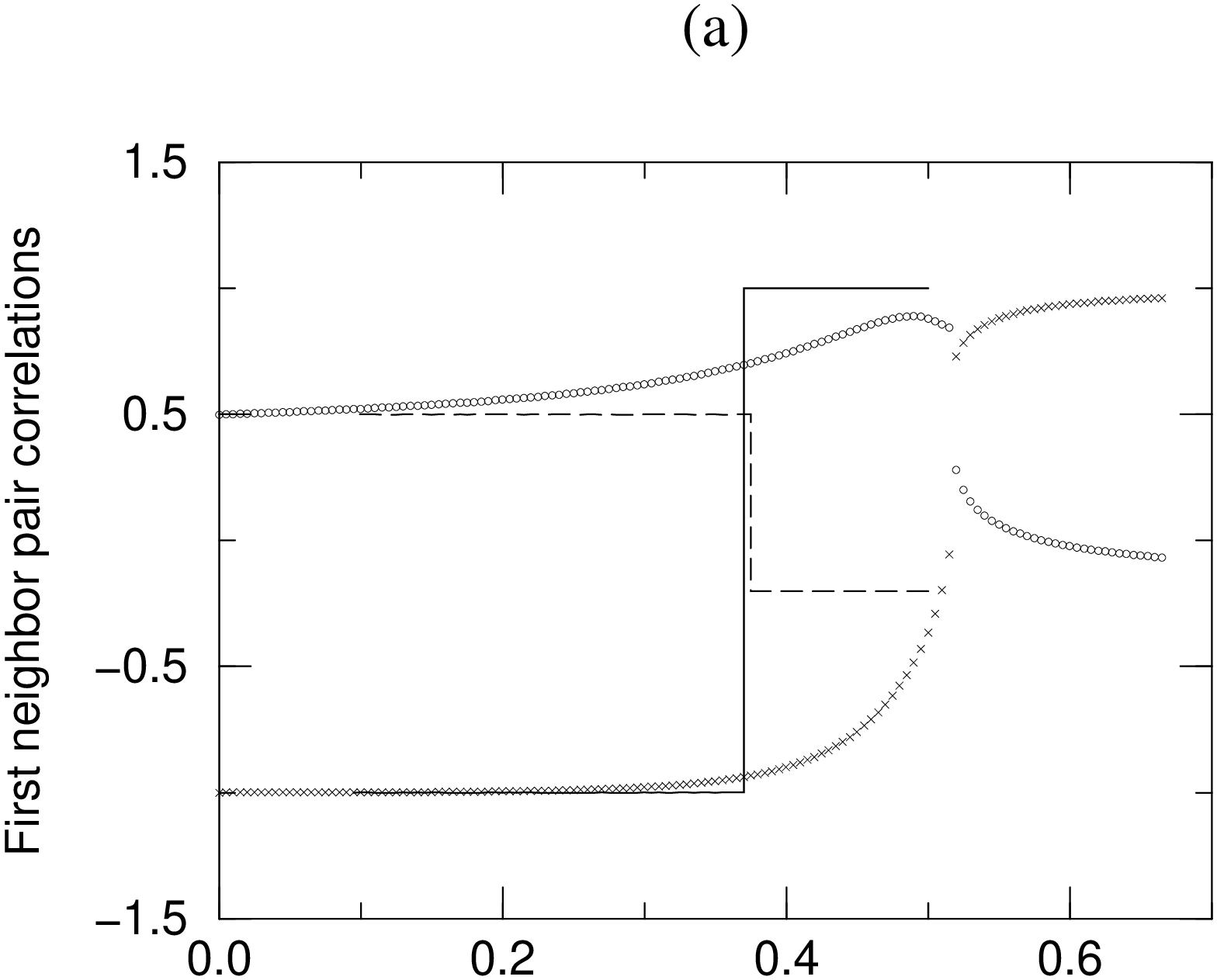}
\end{figure}
\begin{figure}[htbp]
\epsfxsize=7truecm
\epsfbox{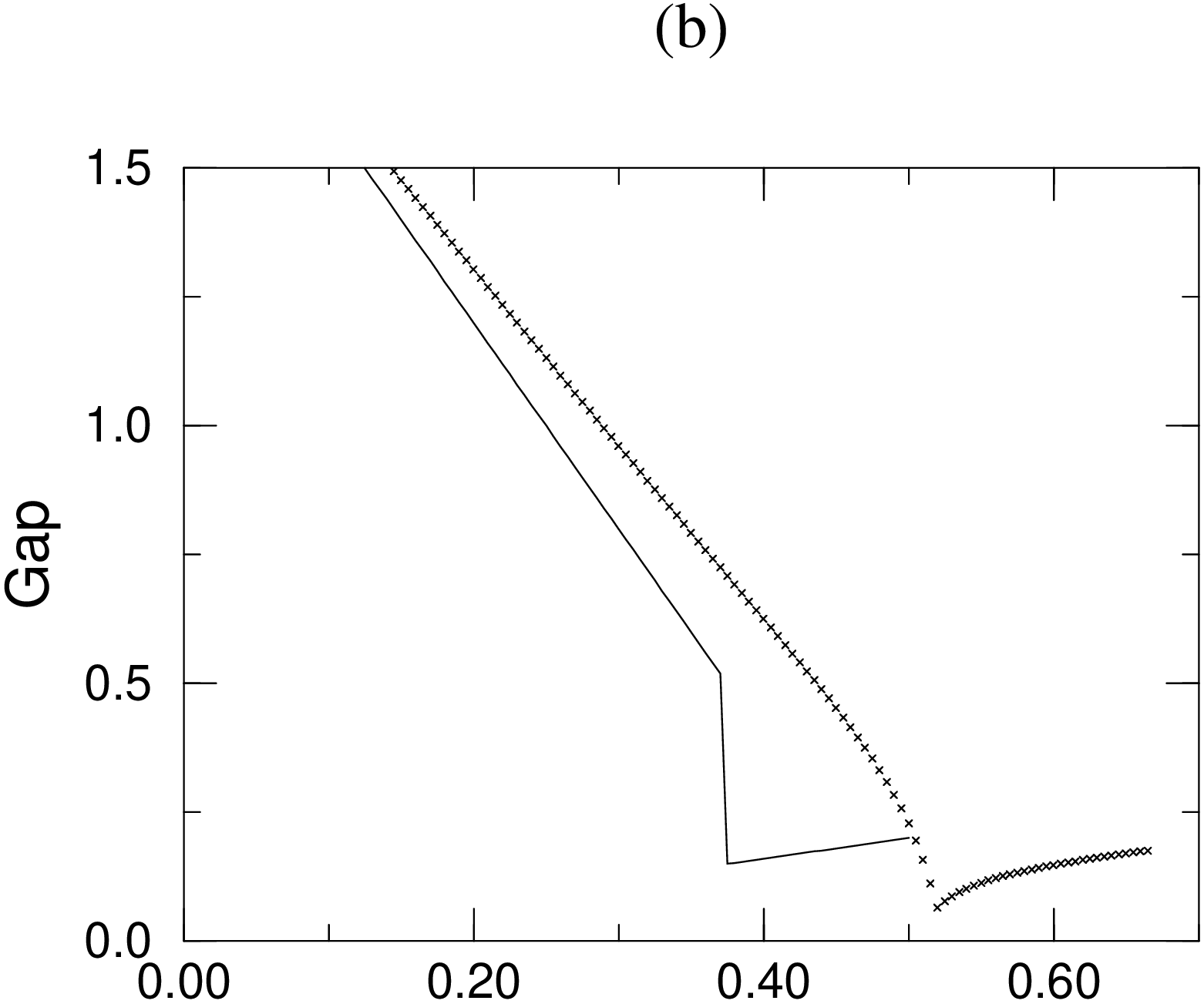}
\end{figure}
\begin{figure}
\epsfxsize=7truecm
\epsfbox{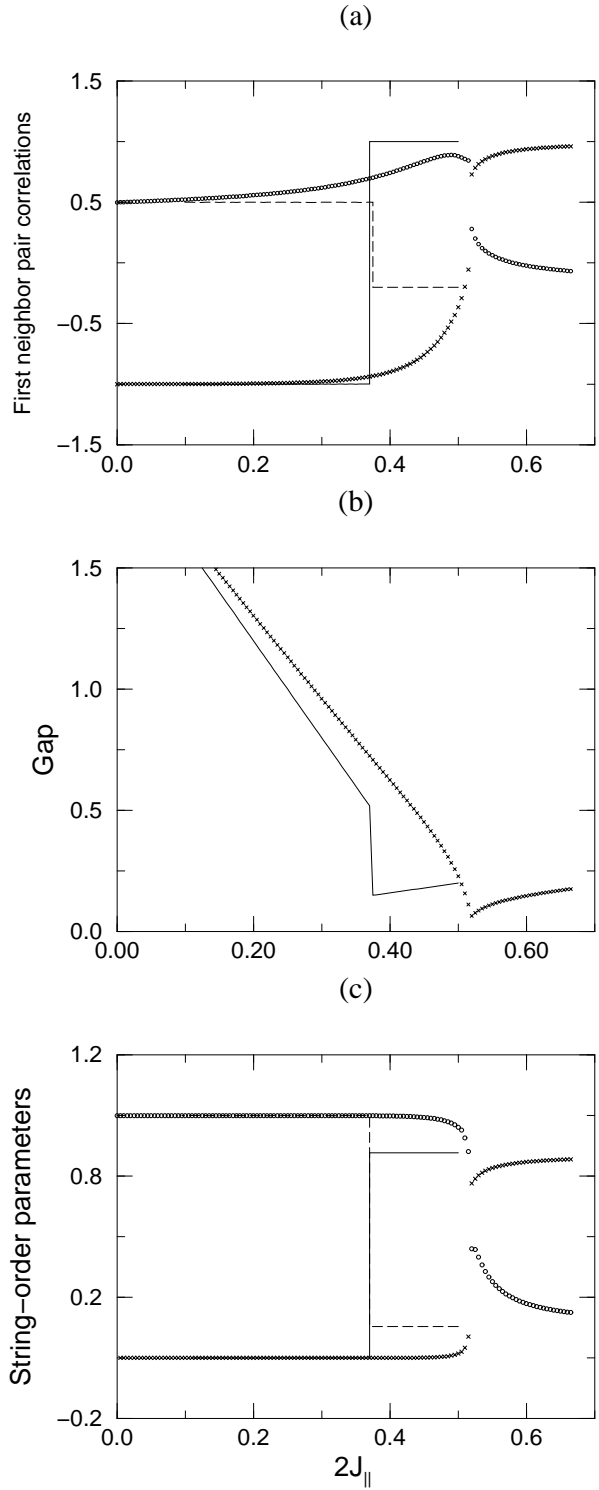}
\caption{First order transition between Haldane phases with different 
topological order. (a) Pair correlation $C_\perp$ along ZW (crosses) and 
ZB (full line); pair correlation $C_{X1}$ along ZW (circles) and ZB (dashed
line). (b) Gap along ZW (crosses) and ZB (full line). (c)  String order
parameter $\tilde {\cal O}_X$ along ZW (crosses) and ZB (full line); 
string order parameter  $\tilde {\cal O}_\perp$ along ZW (circles) and 
ZB (dashed line).}
\end{figure}
\begin{figure}[htbp]
\epsfxsize=7truecm
\epsfbox{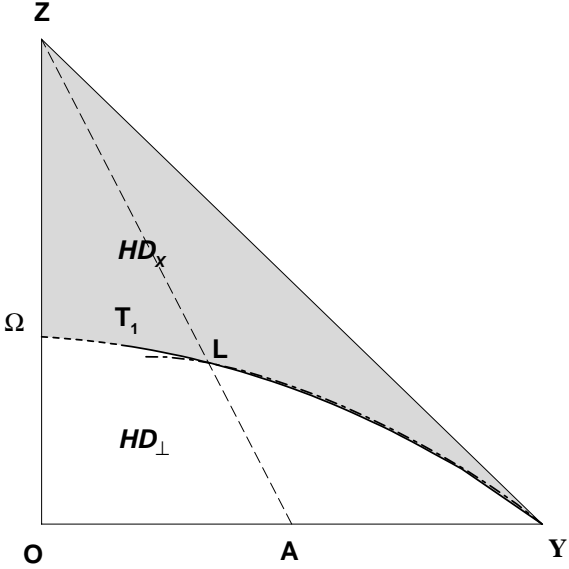}
\caption{The transition line between the two Haldane phases with different
topological order in the $J_\Vert=0$ plane. L represents a simple 
spin-$1\over 2$ antiferromagnetic Heisenberg chain and Y two decoupled chains.
The heavy line YLT$_1$ is critical with zero gap. On T$_1$L (dotted line), 
the transition is first order. The dash-dotted line is a simple approximation
to the critical line. }
\end{figure}
With our nontrivial energy scaling [Eq.~(3)], it corresponds to the real root
of the third-order equation:
\begin{equation}
J_\perp^3-{\cal S}(1-J_\perp)^2-{\cal S}^3+2{\cal S}^2(1-J_\perp)=0
\end{equation}
We have determined through DMRG the transition between the two Haldane phases
in this plane (see Fig.~10). From X to L, the transition line follows 
approximately the analytic form previously proposed but strongly differ
beyond L. The critical line extends beyond L up to a point $\rm T_1$
that we have determined through the same method as that used for the
``Zig-Zag'' model (cf. section IV). 
From $\rm T_1$ to $\Omega$, the transition is first order with finite gap.

\smallskip
\noindent {\it c. The transition surface}
\smallskip

We have exhibited four transition lines, between the two Haldane
phases with different topological order, in four different
planes of the phase diagram: XAZ, XYZ, OBZ, OYZ. We expect that
all four lines belong to a same transition surface. Using the DMRG algorithm
described in section III, we have investigated this transition surface over
a fine grid in the \{x,y\} plane. As a good compromise, we kept 30 to 60
states in the DMRG scheme and chose a $0.02\times 0.02$ \{x,y\} grid. 
\begin{figure}[htbp]
\epsfxsize=7truecm
\epsfbox{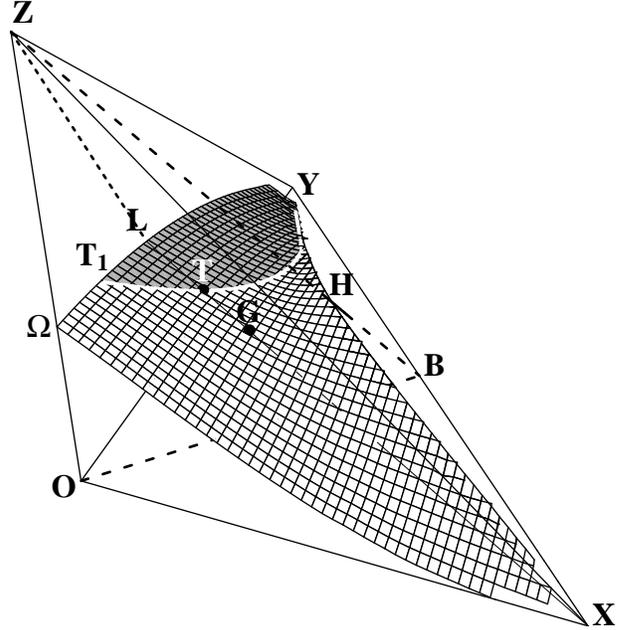}
\caption{The transition surface separating the two gapped Haldane phases with
different topological order in the [+,+,+] sector. 
Each point of the grid has been obtained through DMRG, keeping
30 states. The dark part of the surface is critical with vanishing gap and 
string-order parameters. The clear part corresponds to a first-order
transition with finite gap and degenerate ground state. 
The white line which separates both parts of the
surface thus appears as a tricritical line. 
The cuts of the surface through the planes XALZ, OBHZ, XYZ, and OYZ have been
represented in Figs. 6, 7, 8 and 10 respectively. The straight line LTGX,
joining L (one simple Heisenberg chain) to X (two decoupled Heisenberg chains)
lies in the surface. The Majumdar-Ghosh point G corresponds to the 
intersection of this line with the OBHZ plane.
Near X, our DMRG results are
consistent with the exact result obtained from field theory (see text), i.e.
the transition surface is tangent to the plane 
$J_\perp={\cal S}=J_{X1}+J_{X2}$.}
\end{figure}

The results are shown in Fig. 11. The cuts of this surface through the planes
XALZ, OBHZ, XYZ and OYZ have already been represented in Figs. 6, 7, 8 and
10 respectively. The straight line joining L (simple antiferromagnetic
spin-$1\over 2$ Heisenberg chain) to X (two decoupled Heisenberg chains)
is a symmetry line in the AXZ plane (cf. section IV) and lies in the
transition surface. We used it for an accuracy test on our DMRG algorithm.
For the points of our \{x,y\} grid which lie on the 
projection of this line in the XOY plane, we have plotted in
Fig.~12 the difference
$|z_c^{DMRG}-z_c^{Exact}|$ between the 
approximate coordinate $z_c^{DMRG}$ of the surface obtained through DMRG
and the exact value  $z_c^{Exact}$ corresponding to LX.

\begin{figure}[htbp]
\epsfxsize=7truecm
\epsfbox{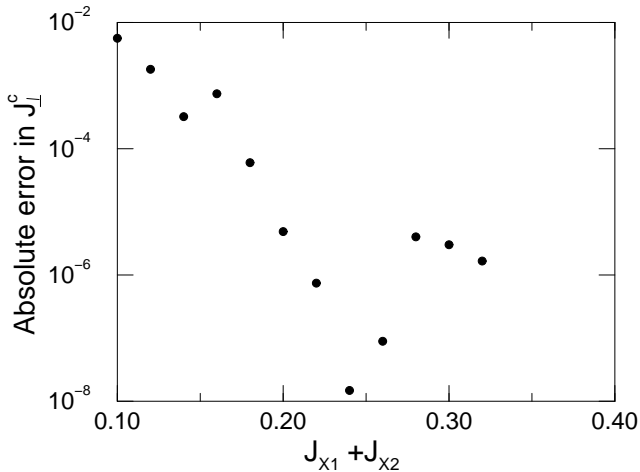}
\caption{A test of accuracy of our DMRG scheme, with only 30 states kept, 
along the exact transition line LX. The accuracy is best (7 digits) near the
Majumdar-Ghosh point G. It decreases drastically to 3 to 2 digits close to X
(two decoupled Heisenberg chains) and also decreases when moving to L
in the critical regime as the second-order transition becomes softer.}
\end{figure}

Except in the neighborhood of X: two decoupled spin-$1\over 2$ 
Heisenberg chains, where a much larger number of states $\approx 30^2$ should 
be retained in the DMRG scheme to obtain precise results, we obtain an 
accuracy which is better than three to four digits on $z_c$. The accuracy is 
the best (7 digits) near the Majumdar-Gosh point G.
It decreases drastically to  $3\sim 2$ digits close to X
(two decoupled Heisenberg chains) and is of about 5 digits near L, 
in the critical regime.

Field-theoretical results\cite{kim} prove that in both planes XYZ and
XALZ the transition line, near X,
 is first order and correspond to ${\cal S}=J_\perp$.
In agreement with these exact results, our numerical data
indicate that the transition surface is tangent at X to the
plane ${\cal S}=J_\perp$.

Knowing this transition surface, we now determine which part of it is
critical, with zero gap. We already know two intersecting critical lines 
LT$_1$ and YLT. We expect that they lie in a same critical part of the
surface which is limited by a line of tricritical points 
({\it ``tricritical line''}) passing through T and T$_1$.
To determine this tricritical line, we choose the same methods as these
previously used for the ``Zig-Zag chain'' model\cite{emery,eggert} (see
section IV).

It is proved from
field theory that, when umklapp interactions become relevant variables, 
simultaneously the
spectrum develops a gap between the S=0 ground state and the first excited
S=1 states and also  the degeneracy of the first S=0 excited state and
lowest S=1 state is split. 
Hence in the thermodynamical limit, the difference $\delta$
between the energies of these two lowest excited states is zero at the
critical value for exchange parameters where the gap vanishes. 
As a simple illustration, let us consider the curved line $\Omega$LY (Figs. 10 
and 11) corresponding to the cut of the transition surface through the
OYZ plane. 

\begin{figure}[htbp]
\epsfxsize=7truecm
\epsfbox{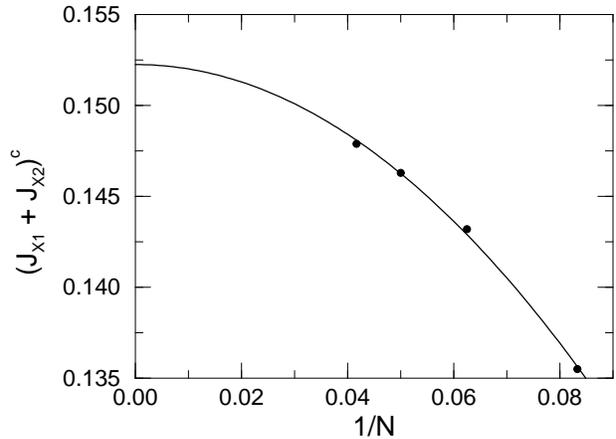}
\caption{Critical value of ${\cal S}=J_{X1}+J_{X2}$ for which the difference
$\delta$ between the two lowest excited states vanishes, as a function of
1/N, for finite chains with periodic boundary conditions. We follow the
transition line $\Omega$LY determined through DMRG 
in the OYZ plane (Fig. 10). A quadratic
extrapolation gives $(J_{X1}+J_{X2})^c\approx 0.152$ in the thermodynamic
limit, fixing the position of T$_1$ on the transition line.
}
\end{figure}

Moving along this curve, we use Lancz\"os diagonalisation on
finite systems of length $L=N/2$, with periodic boundary conditions, to
calculate the value ${\cal S}_c(N)$ of the parameter ${\cal S}=J_{X1}+J_{X2}$
at which $\delta$ vanishes.

\begin{figure}[htbp]
\epsfxsize=7truecm
\epsfbox{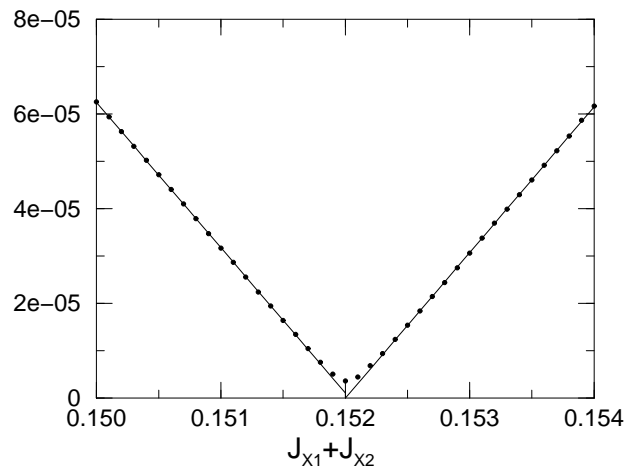}
\caption{Mean square deviation with respect to a $1/N^3$ law of the difference
between the two first excited levels of a finite ladder with periodic boundary
conditions. We follow the cut of the transition surface
through the plane $2J_\Vert =0$  determined through DMRG.}
\end{figure}
 The results are represented in Fig. 13 as a 
function of $1/N$. A quadratic extrapolation gives ${\cal S}_c(N\to\infty)
\approx 0.152$ in the thermodynamic limit. 

A much better accuracy (six digits) on the position of T on LX, for the
``Zig-Zag chain'' model has been 
obtained recently by Eggert\cite{eggert}.
He used the field-theoretical
result that, on LX at the critical point T, the marginal operator 
vanishes and the energy difference $\delta (N)$ between the two first excited 
levels is exactly proportional to $1/N^3$.
 Although we have not been able
to generalize this field-theoretical result for every point on the tricritical
line, we have found numerically that except in the vicinity of the boundary
plane XYZ, this $1/N^3$ law works remarkably well for a large part of the
tricritical line. And we have used it to improve our accuracy.

\begin{figure}[htbp]
\epsfxsize=7truecm
\epsfbox{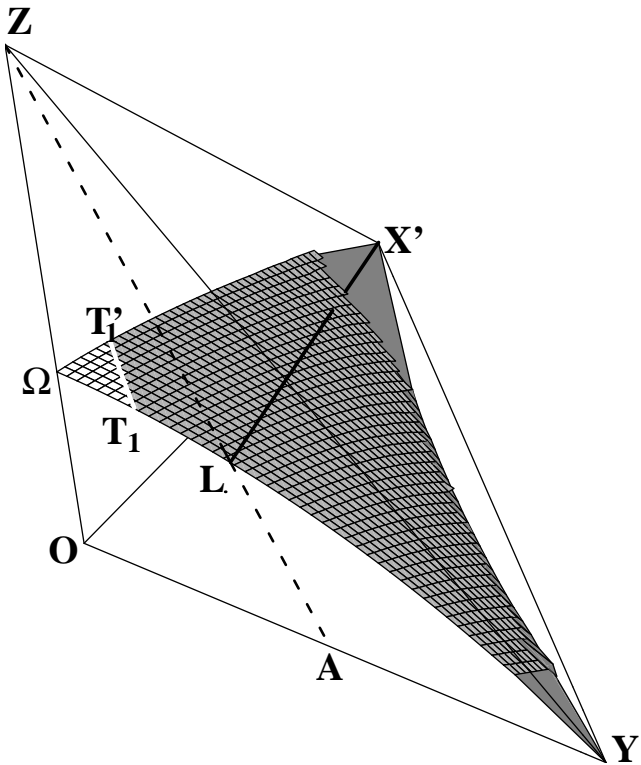}
\caption{The transition surface between the two Haldane phases with different
topological order in the $[-,+,+]$ sector. The dark part of the surface
corresponds to the zero-gap critical part. All points of the mesh are
DMRG results. Our DMRG result are not accurate in the vicinity of 
ferromagnetic (X') or antiferromagnetic (Y) decoupled chains. The grey areas
are simple extrapolations. The critical part of the surface is bounded through
the tricritical T$_1$T'$_1$ white line. On  T$_1$T'$_1\Omega$ the transition
is first order with two degenerate ground-states and finite gap. The cut
by the plane ALZX' corresponds to the ``Zig-Zag chain'' model (cf. Fig. 6 in
section IV). The straight line LX' lies in the critical surface. 
}
\end{figure}

Figure 14 illustrates around the critical value ${\cal S}=0.152$ found
by the previous method (Fig. 13) the mean square deviation of $\delta(N)$
with respect to a $1/N^3$ law, when we follow the transition line determined
through DMRG in the OYZ plane. The results are based on Lancz\"os 
diagonalizations on finite systems with N=16, 20 and 24 sites. The
critical value thus obtained  ${\cal S}=0.15201$ is improved up to five
digits. This corresponds to the accuracy on the position of the
transition line obtained through DMRG (see Fig. 12).

Taking successive cuts of the transition surface through different vertical
planes at $J_\Vert=C^{st.}$ or ${\cal S}=C^{st.}$ 
and using the same methods, we have determined  the 
whole tricritical line T$_1$TY representing the boundary of the critical part
of the transition surface (heavy white line in Fig. 11). Near Y, this line
is close to the boundary of the surface of transition in the XYZ plane. We
do know however that it does not intersect it: 
exact field-theoretical results\cite{kim}
prove that in the XYZ plane the transition is first order near Y. 
\subsubsection{The $[-,+,+]$ sector}
The transition surface and its zero-gap critical part is determined
though the same methods as used in the previous subsection. The results
are shown in Fig. 15. The dark part of the surface is critical with zero
gap. It does not intersect the X'YZ boundary face and extends
further in the $[-,+,-]$ sector. It is bounded through the tricritical line
T$_1$T'$_1$. On the left of this line ( T$_1$T'$_1\Omega$) part, the
transition is first order with degenerate ground state and finite gap.
The cut through the ALZX' plane corresponds to the ``Zig-Zag chain'' model
(see Fig. 6 in section IV). The LX' straight line is a symmetry line in
that plane and it lies in the critical surface.

\subsubsection{The $[-,-,+]$ sector}
This sector is particularly interesting since the two Haldane phases
with different topological order compete here with the ferromagnetic phase.

\smallskip
\noindent {\it a. The ferromagnetic volume}

\smallskip
From section IV~A, the A'X'Z plane corresponding to the ``Zig-Zag chain''
model $(J_{X1}=0)$  splits the ferromagnetic boundary in two parts. On the
$J_{X1}>0$ side the exact boundary of the ferromagnetic domain is the
surface determined by Eq. (20), taking into
account our energy scaling [Eq. (3)]. On the other side $(J_{X1}<0)$,
we have calculated the ferromagnetic boundary with an accuracy of
three to four digits through DMRG (see Fig. 16).
The straight line
 segment CD corresponding to the intersection with the OB'Z plane
(cf. Fig. 7) lies in this boundary surface.
Near $X'$, we have two weakly coupled ferromagnetic chains. At first order
in perturbation theory, our Hamiltonian can be written as:
\begin{equation}
{ H}_{eff}=LJ_\Vert + (J_\perp +{\cal S}){\bf S_T^a\cdot S_T^b}/2L
\end{equation}
where
$$
{\bf S_T^a}=\sum_{i=1}^L{\bf S_{2i}} \quad {\rm and} \quad
{\bf S_T^b}=\sum_{i=0}^{L-1}{\bf S_{2i+1}}
$$
represent the total spins of the two coupled ferromagnetic chain.
We deduce that the plane $(J_\perp +{\cal S}=0)$ is tangent at X' to the
ferromagnetic surface.

\begin{figure}[htbp]
\epsfxsize=7truecm
\epsfbox{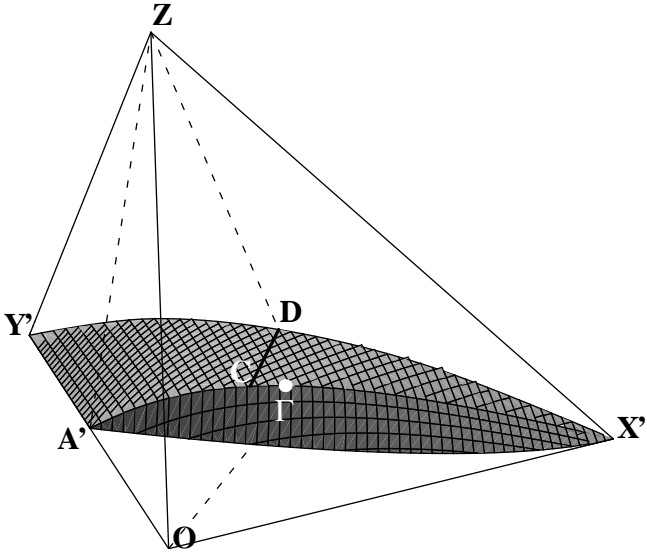}
\caption{The surface limiting the ferromagnetic phase in the $[-,-,+]$ sector.
The line A'C$\Gamma$X' at the intersection of this surface with the A'X'Z
$(J_{X1}=0)$ plane corresponds to the exactly soluble DKO model.
The front part of the surface corresponding to $(J_{X1}>0)$ is also known 
exactly [Eqs.~(3,~20)]. The back part of the surface corresponding to
$(J_{X1}<0)$ has been determined through DMRG. The straight line CD 
corresponding to its intersection with the OBZ plane (see Fig. 7) is exact.
The line ACX' corresponding to the intersection of the surface with the
A'X'Z plane is singular. From A' to $\Gamma$, our DMRG results show a
discontinuity in the gradient to the surface. From   $\Gamma$ to X, there
are singularities in higher derivatives. The Fig. 18 show that this line
corresponds to the intersection
of the ferromagnetic surface with the transition surface between the two
Haldane phases with different topological order.  $\Gamma$ is at the
crossing of the A'CX' line with the tricritical line limiting the critical
part of this latter transition surface.}
\end{figure}
\begin{figure}[htbp]
\epsfxsize=7truecm
\epsfbox{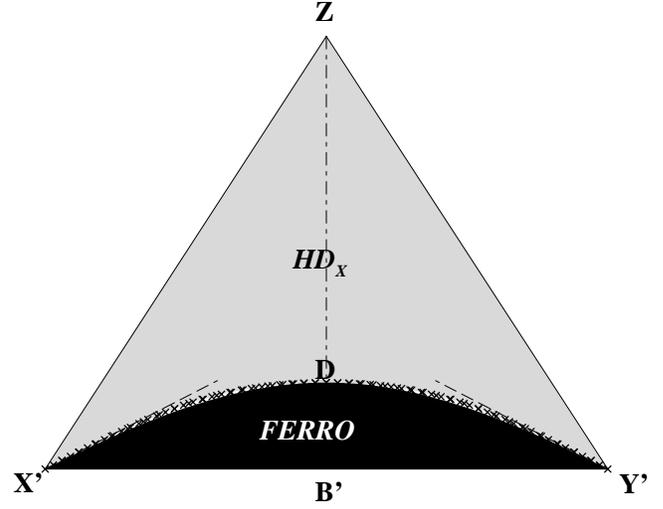}
\caption{The first order transition between the $HD_X$ and ferromagnetic phase
in the X'Y'Z face. The dashed lines are the cuts through the exact
tangent planes to the ferromagnetic boundary at X' and Y' 
(two independent ferromagnetic chains). The dots are our DMRG results.
the boundary of the black area represents a simple inner bound to the
ferromagnetic region obtained through [Eqs.~(3,~21)].
}
\end{figure} 

Similarly, the whole segment A'Y' of the OY' axis corresponds to two
decoupled identical ferromagnetic chains with alternating $(J_{X1},~J_{X2})$
ferromagnetic bonds. The segment A'Y' lies in the ferromagnetic
boundary surface and near A'Y', at first order in perturbation theory
the Hamiltonian can be expressed as:
\begin{equation}
{ H}_{eff}=L{\cal S} + (2J_\Vert+J_\perp){\bf S_T^a\cdot S_T^b}/2L
\end{equation}
where
$$
{\bf S_T^a}=\sum_{i=1}^{L/2}({\bf S_{4i}+S_{4i-3}}) \quad {\rm and} \quad
{\bf S_T^b}=\sum_{i=0}^{L-1}({\bf S_{4i-1}+S_{4i-2}})
$$
and the plane $(2J_\Vert+J_\perp)=0$ is tangent to the ferromagnetic
boundary all along the segment A'Y'.

 Although the inner envelope to this boundary surface
determined by Eqs.~(3,~21) is not exact, it is everywhere an excellent
approximation (see the comparison in the X'Y'Z plane represented in Fig. 17).
This is due to the fact that three different exact transition lines:
A'Y', $CD$ and A'CXX' belong to this envelope and that the planes
previously considered are also tangent to this envelope.

From our DMRG results, it is clear that the line A'C$\Gamma$X', at the 
intersection of the A'X'Z plane (``Zig-Zag chain'' model) with the 
ferromagnetic surface and corresponding to DKO exact model is 
singular. On this line, there are discontinuities in the gradient to
the surface from A to $\Gamma$, a point whose position will be
given below. From $\Gamma$ to X', there are singularities in higher-order 
derivatives. 

\begin{figure}
\epsfxsize=7truecm
\epsfbox{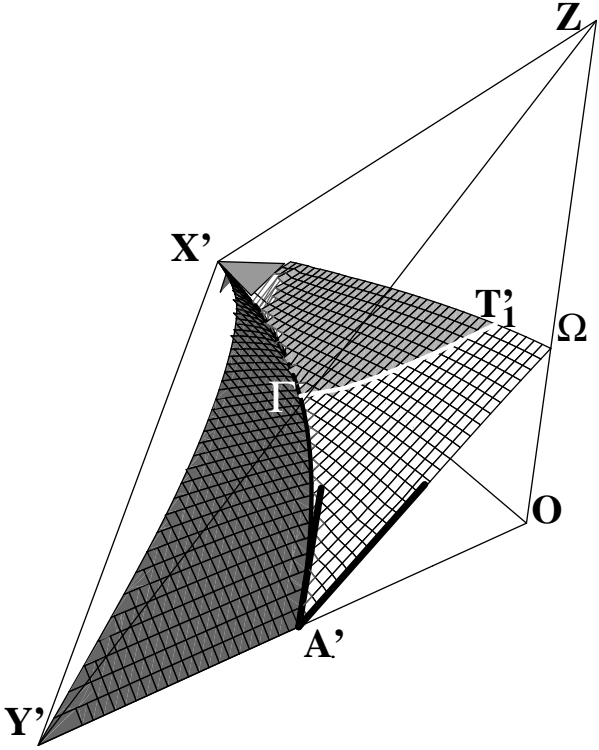}
\caption{Transition surfaces in the $[-,-,+]$ sector. The dark-grey surface
A'X'Y' is the boundary of the ferromagnetic phase and corresponds to a 
first-order transition between the Haldane $HD_X$ and the ferromagnetic
phase. The other surface corresponds to the transition between the two
Haldane phases $HD_X$ and $HD_\perp$ with different topological order. The
dark part of this surface is critical with zero gap and string order 
parameters. On the clear part, the transition is first order and the gap is 
finite. This surface abuts the ferromagnetic boundary along a line
A'$\Gamma$X' which lies in the AX'Z plane
$(J_{X1}=0)$ and corresponds corresponds to the exactly soluble DKO model.
The white line T'$_1\Gamma$ limiting the critical zero gap surface ends
on the DKO line at $\Gamma$. The part A'$\Gamma$ of the
DKO line is ``Triple Line'' while $\Gamma$X' is a ``Critical-End
Line''.
 The two heavy lines starting from A' represent the plane
 $J_{X1}=0.7444(2J_\Vert + J_\perp)$, where a first order transition
between two Haldane phases with different topological order has been found
by Kolezhuk et al. in the frustrated spin-1 chain. This plane is tangent
to our transition surface. }
\end{figure}

\smallskip
\noindent {\it b. Triple line, critical-end line}

\smallskip
We have calculated the transition surface between the two Haldane phases 
with different topological order in the same way as in the other sectors
(see Fig. 18)

To our numerical accuracy, we observe that the meeting of this 
surface with the ferromagnetic boundary occurs in the X'A'Z plane and
corresponds to the DKO exactly-soluble model. 
The X'T'$_1\Gamma$ part of this surface is critical with vanishing gap and
string-order parameters. On the other part A'$\Gamma$T'$_1$ the transition
is first order and the antiferromagnetic ground-state is doubly degenerated.

The tricritical line T'$_1\Gamma$ limiting the critical zero-gap surface ends
on the DKO line at a particular point $\Gamma$:(-0.35,-0.2518,0.1464).
According to the usual terminology
concerning three dimensional phase diagrams, the part A'$\Gamma$ of
the DKO line is a line of triple points (i.e. a {\it ``Triple Line''})
while $\Gamma$X' is a line of critical-end points (i.e. a {\it ``Critical-End
Line''}).

Our numerical results concerning the coincidence of the meeting of the
two surfaces with the DKO line are enforced by the
following theoretical argument:

\begin{itemize}
\item
In the region  with $J_{X_1}<0$ all interactions except $J_\perp$ are 
ferromagnetic. Consequently singlet pairing (``valence bond'') can only occur
along the rungs, while triplet pairing can take place along diagonals.
Consequently only the  $HD_X$ Haldane phase can appear. We conclude that the
transition surface between the two Haldane phases cannot cross the 
$(J_{X1}=0)$ A'X'Z plane. It only exists in the  $J_{X_1}>0$ side.

\item
In the  $J_{X_1}>0$ side, the ferromagnetic boundary is exactly known
and corresponds to the simple Equation 20. This surface has no singularity
in its gradient or higher order derivatives. The meeting of the two surface
should correspond to a singularity on the ferromagnetic surface: discontinuity
in the gradient for the first-order part of the  surface, or 
singularities in higher order derivative for the critical part. 
Consequently the transition surface between the two Haldane phases do not
touch the  $J_{X_1}>0$ part of the ferromagnetic boundary.

\item
From the two previous remarks, we conclude that if the transition surface
between the two Haldane phases meets the ferromagnetic surface, it can only
be in the A'X'Z plane ($J_{X_1}>0$)  i.e. at the DKO line.

\end{itemize}

\smallskip
\noindent {\it c. The neighborhood of the frustrated spin-1 Heisenberg chain}

\smallskip
In section II~A we have shown that the neighborhood of A' can be described
through the frustrated spin-1 chain model (Eq. 5) with effective first-neighbor
 $J_1=2J_\Vert+J_\perp$ and second-neighbor $J_2=J_{X1}$ interactions.
This model has recently been studied by Kolezhuk et al\cite{kolezhuk}
through DMRG. At a critical ratio $r_c=J_2/J_1 \approx 0.7444$, there is
a sharp discontinuity in the order parameter ${\cal O}_X$ suggesting
a first order transition. This critical ratio corresponds here to the
plane $J_{X1}=r_c(2J_\Vert + J_\perp)$. We have represented this plane
in Fig. 18. It is tangent in A' to our transition surface. Consequently
our diagram also contains the results on the frustrated spin-1
Heisenberg chain, and the transition observed by Kolezhuk et al. corresponds
to a first-order transition between the $HD_X$ and $HD_\perp$ phases.

\subsubsection{The $[+,-,+]$ sector} 
The transition surface between the two Haldane phases in the last of the 
four sectors with $J_\perp>0$ is represented in Fig. 19.
Except in the thin tip close to X we have found no evidence for a zero-gap
critical part. The poor accuracy of our DMRG results near X does not
allow to be more quantitative.
Again, near A', our calculated surface is tangent to the plane
$J_{X1}=0.7444(2J_\Vert + J_\perp)$ corresponding to the results of Kholezhuk
et al.\cite{kolezhuk} (see above)

An overall view of the four previous sectors corresponding to $J_\perp>0$
is shown in Fig. 20.

\subsubsection{The $[-,-,-]$ sector with  with $J_\Vert$, $J_\perp$ and
               ${\cal S}$ negative}
In this sector the phase transitions are between the ferromagnetic and the
$HD_\perp$ Haldane states. The corresponding surface is known exactly and
corresponds to Eqs.~(3,~20). It is shown in Fig.~21.
\begin{figure}
\epsfxsize=7truecm
\epsfbox{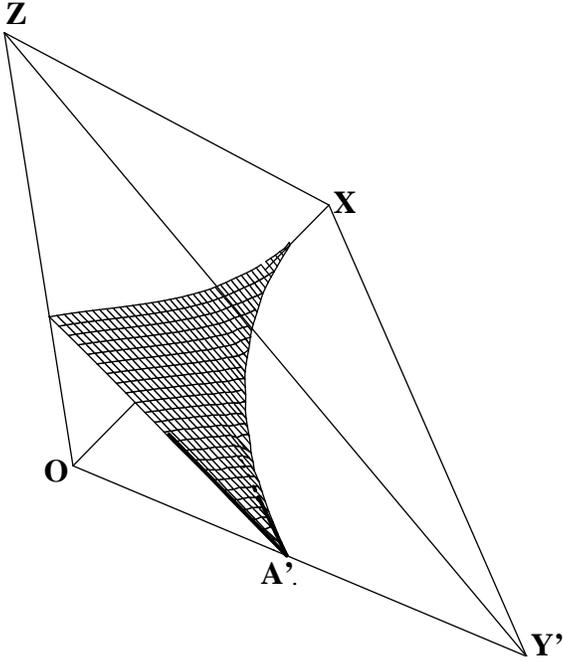}
\caption{The first order transition surface between the two Haldane phases
$HD_X$ and $HD_\perp$ in the $[+,-,+]$ sector.
 The two heavy lines starting from A' represent the plane
 $J_{X1}=0.7444(2J_\Vert + J_\perp)$, where a first order transition
between two Haldane phases with different topological order has been found
by Kolezhuk et al. in the frustrated spin-1 chain. This plane is tangent
to our transition surface}
\end{figure}
\begin{figure}
\epsfxsize=7truecm
\epsfbox{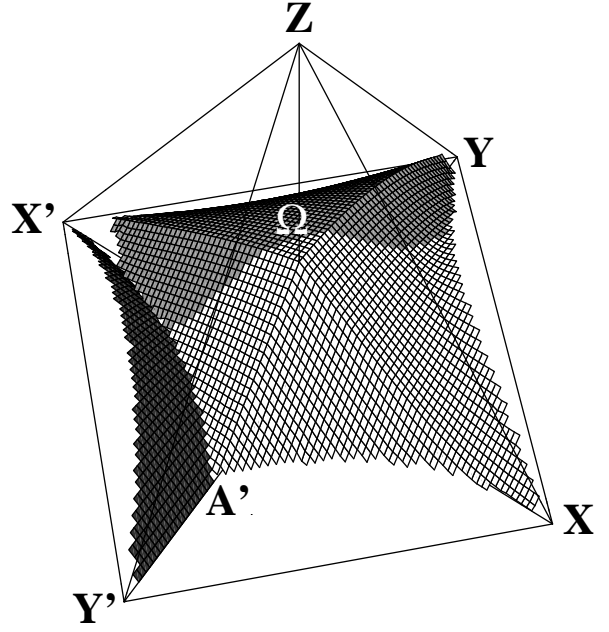}
\caption{An overall view of the transition surface in the upper half space.
The black surface A'X'Y' represents the boundary of the ferromagnetic
phase. The other surface represents the transition between the two Haldane
phases $HD_X$ and $HD_\perp$. The dark part is critical with zero gap,
the clear part corresponds to a first-order transition
with finite gap. Singularities on the lines $\Omega$A, $\Omega$X, $\Omega$Y,  
and $\Omega$X' are not physical: they are related to our normalization 
condition (Eq. 3). }
\end{figure}

\begin{figure}
\epsfxsize=7truecm
\epsfbox{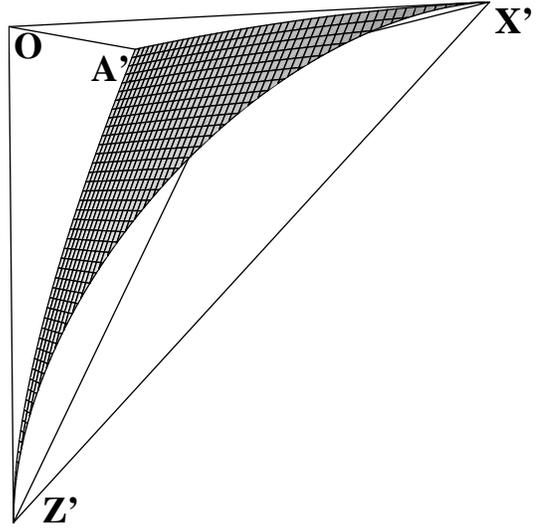}
\caption{The exact limit of the ferromagnetic phase in the [-,-,-] sector}
\end{figure}

\subsubsection{The $[+,-,-]$ sector}

The boundary face XY'Z' of this sector corresponds to the model represented
in Fig.~1~(b) with ferromagnetic equal 
diagonal interactions and antiferromagnetic
$J_\Vert$ interactions. There is an obvious symmetry by exchanging the
parallel and diagonal interaction which is obtained by twisting every other
rung by an angle $\pi$ around the longitudinal axis of the ladder. Hence there
is a one to one correspondence from each point of this face to a point of the
YX'Z' face in the $[-,+,-]$ sector. The two-dimensional phase-diagram of
the XY'Z' face is represented in Fig. 22. The transition line XE' between the
two Haldane phases $HD_X$ and $HD_\perp$ is critical. Near X this critical
line corresponds to $J_\perp={\cal S}$ in agreement with field-theoretical
results\cite{kim}. The critical line remains close to this asymptotic
result except in the neighborhood of the ferromagnetic region, near E'.

\begin{figure}
\epsfxsize=7truecm
\epsfbox{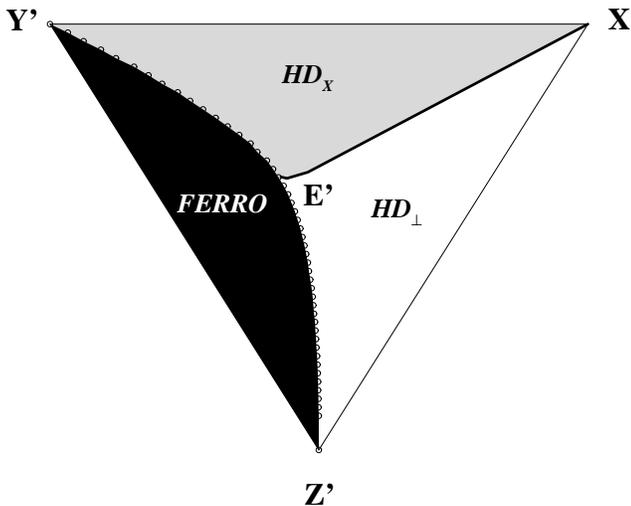}
\caption{The two-dimensional 
phase diagram in the boundary face XY'Z' corresponding to equal
diagonal interactions [Fig.~1~(b)]. The diagonal interactions are ferromagnetic
while $J_\Vert$ is antiferromagnetic. There is a symmetry which exchanges
parallel and diagonal interaction by twisting every other rung by an angle
$\pi$ around the longitudinal axis. Consequently the phase diagram of the
YX'Z' face is identical. The transition line XE' between the two Haldane phases
$HD_X$ and $HD_\perp$ is critical and remains close to its asymptotical
limit at X $(J_\perp={\cal S})$ obtained from field theory, except near its
meeting with the ferromagnetic boundary at E'. }
\end{figure}

The three-dimensional phase-diagram of the $[+,-,-]$ sector is shown in 
Fig.~23. The dark surface represents the ferromagnetic boundary seen
from inside. It
cuts the $(J_{X1}=0)$ A'XZ' plane at the A'C'Z' line corresponding to
the exactly soluble DKO model. The part of this surface which
corresponds to $(J_{X1}>0)$ is known exactly from Eq.~(20). The other part
is calculated through DMRG. The other surface represents the
transition between the two Haldane phases $HD_X$ and $HD_\perp$. The
clear part corresponds to a first order transition with finite gap and
string-order parameter. The other part is critical with vanishing gap and 
string-order parameter. The blank tricritical line bounding this critical
part has been determined through the methods explained above. Its
intersection with the XOY' plane corresponds to $2J_\Vert\approx 0.75$
and the critical surface has a small extension in the previous $[+,-,+]$
sector around X. However our DMRG scheme is inadequate to 
determine the critical surface so close to 
X (two decoupled chains) with reasonable accuracy (see Fig. 19).

The symmetry line XC', in the $(J_{X1}=0)$ A'XZ' plane lies in the transition
surface (see Fig. 6b). C' represents the intersection of the line XL' joining
X (two decoupled antiferromagnetic chains) to L' (one ferromagnetic
chain) with the  DKO line. The surface starts at the $J_\perp=0$ plane
roughly perpendicular to it and it curves sharply around XC' to meet the
boundary plane XY'Z' along XE', nearly tangentially to the 
${\cal S}=J_\perp$ plane. 
It abuts the ferromagnetic boundary along the A'C'E' curve, 
which is represented
from inside the ferromagnetic region by a heavy line in Fig.~23. 

The part
A'C' of this curve lies in the A'XZ' plane and coincides with the DKO curve.
Since it corresponds to the meeting of two first-order transition surfaces,
the curve A'C' is a triple line.
This is justified by the following arguments:

\begin{itemize}
\item
The transition surface between the two Haldane phases crosses the
 $(J_{X1}=0)$ A'XZ' plane at the symmetry line XC'. There is a one to one
correspondence on each side of the XC' line in the A'XZ' plane which maps
the  $HD_X$ to the $HD_\perp$ phase by exchanging rungs and diagonal.
Consequently there is no crossing of the transition surface at any other
place in the A'XZ' plane. The transition surface lies on the same side
$(J_{X1}>0)$ of the A'XC'L' triangle.

\item
On this side of the A'XZ' plane $(J_{X1}>0)$, the ferromagnetic
boundary is exactly known [Eq. (20)]. It has no singularity, which means
that the transition surface between the two Haldane phases does not
connect to it.

\item
From the two previous remarks, we conclude that the transition surface
between the two Haldane phases abuts the ferromagnetic boundary along the curve
AC' in the A'XZ' plane $(J_{X1}=0)$

\end{itemize}

\begin{figure}
\epsfxsize=7truecm
\epsfbox{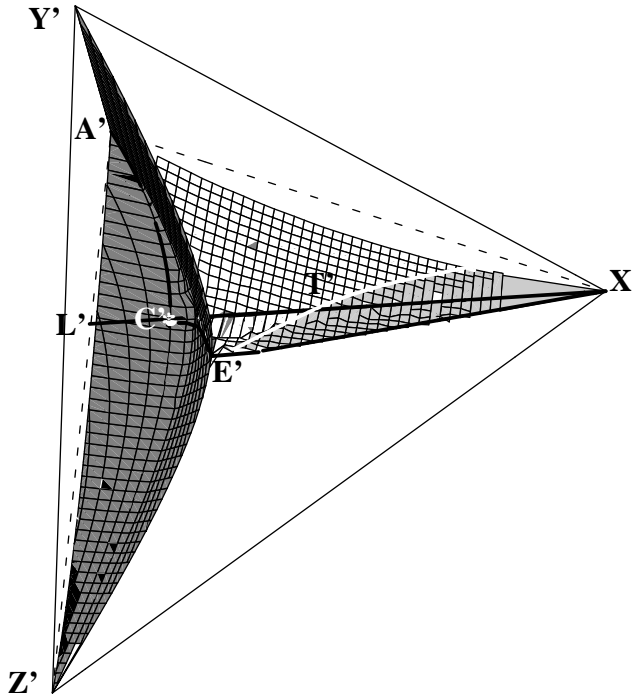}
\caption{The 3D phase diagram in the $[+,-,-]$ sector. The cut through the
boundary face XY'Z' is shown in Fig. 22 and the cut through the plane
A'XZ' corresponding to the ``Zig-Zag chain'' model has already been studied
in Fig.~6~(b). The dark gray surface corresponds to the boundary of the
ferromagnetic phase seen from inside. 
For $J_{X1}>0$, the ferromagnetic boundary is given
exactly through Eq. (20). The other side $J_{X1}<0$ is calculated through
DMRG. The other surface represents the transition between the two Haldane
phases $HD_X$ and $HD_\perp$. The clear part corresponds to a first-order
transition with finite gap and string-order parameters. The dark part
is critical with vanishing gap and order parameter. The tricritical white
line represents the boundary of the critical part of the surface. 
This surface abuts the ferromagnetic boundary along the curve A'C'E',
which is represented from inside the ferromagnetic boundary as a heavy
line. The part A'C' of this curve lies in the A'XZ' plane $(J_\perp=0)$
and coincides with the DKO line. It is a triple line.}
\end{figure}

The other part C'E' of this curve is clearly out of the A'XZ' plane. The
meeting of this curve with the tricritical line is numerically found very
close to E'. We have not been able however to prove that it coincides
exactly with E'. We conclude that at least part of the curve C'E' is a
triple line.

\subsubsection{The $[+,+,-]$ sector}

No phase transition has been found in this sector. The ground state is 
everywhere the $HD_\perp$ phase.

\subsubsection{The $[-,+,-]$ sector}

\begin{figure}
\epsfxsize=7truecm
\epsfbox{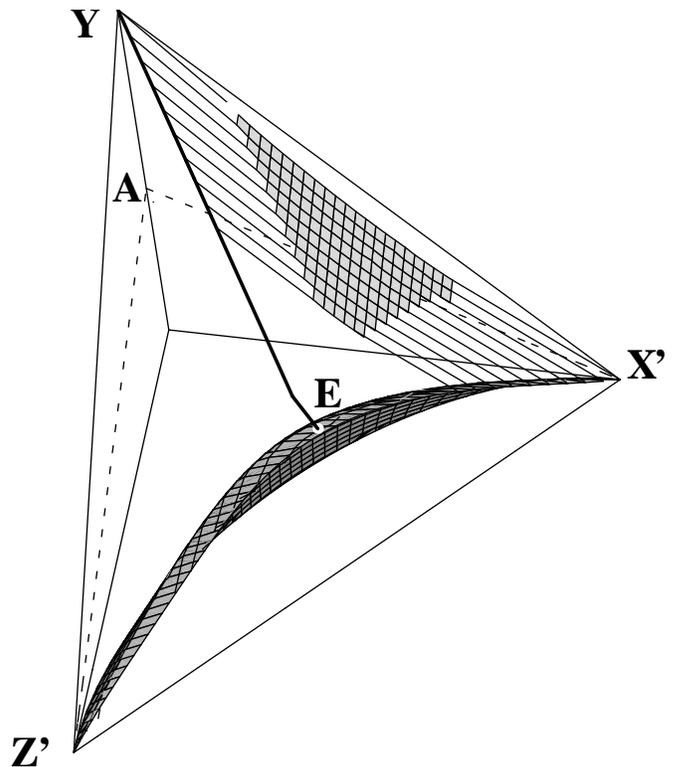}
\caption{The 3D phase diagram in the $[-,+,-]$ sector. The boundary face
YX'Z' is obtained from the symmetrical diagram represented in Fig. 22 by
changing X to Y and Y' to X'. The dark surface represents the boundary of
the ferromagnetic phase. Its intersection with the $(J_{X2}=0)$ AX'Z' plane
corresponds to the DKO exactly-soluble model. The $(J_{X2}>0)$ part of
this surface is known exactly [Eq.~(20)]. The other part is calculated
through DMRG. A part of the critical surface (clear-gray grid) corresponding
to a second order transition between the two Haldane phases
phases $HD_X$ and $HD_\perp$ has been determined through DMRG. It is very
close to the boundary face X'YZ' but cuts it only on YE. 
A rough extrapolation of this surface (array of thin lines) is also
shown; it indicates that the intersection of the critical surface with
the ferromagnetic boundary is close to the X'YZ' face }
\end{figure}

The three-dimensional phase diagram of the last $[-,+,-]$ sector is
shown in Fig. 24. The ferromagnetic boundary (dark surface) is obtained
in the same way as in other sectors: the part corresponding to
$(J_{X2}<0)$ is known exactly from Eq. (20).
 The complementary part is calculated
through DMRG. 

The cut by the X'YZ' face is obtained from Fig. 22 (XY'Z' face) through
the symmetry which exchanges parallel and diagonal interactions.
The transition surface between the two Haldane phases  $HD_X$ and $HD_\perp$
lies everywhere close to this boundary plane and cuts it on YE.

 Only a
part of this surface (clear gray grid in Fig. 24) has been obtained,
using our DMRG scheme, with reasonable accuracy. It is critical with
vanishing gap and string order. The array of thin lines in Fig. 24 is a rough
extrapolation. It indicates that the intersection of this surface with
the ferromagnetic boundary lies also close to the boundary face X'YZ'.

\section{Conclusion}

We have presented an overall view on the three-dimensional
phase diagram of the frustrated
two-leg ladder model. This model contains, as particular cases, the
spin-1 Heisenberg chain with first and second nearest-neighbor interactions,
the spin-$1\over 2$  ``Zig-Zag chain'' with dimerization and frustration,
the usual ladder and a variety of previously studied ladder models that we have
briefly reviewed. We have proven that only three different phases occur:
the ferromagnetic phase and two gapped Haldane phases with different
symmetry and different long-range topological order. In a three dimensional
phase diagram, these three phases are separated through transition surfaces.
Some parts of the transition surface separating the two gapped Haldane
phases are critical with vanishing gap and string-order parameters; the
complementary parts corresponds to a first order transition with finite
gap and two degenerate S=0 ground-states. In the 
two dimensional plane cuts of the 
three-dimensional phase-diagram which correspond
to the ``Zig-Zag chain model'', there is a one to one mapping
of these two Haldane phases through a simple symmetry which exchanges
rung and diagonal interactions. Consequently, although the topology of
these two Haldane phases differ, the critical behavior concerning their
thermodynamic properties near the critical surface are the same and they
belong to the same class of universality.
Part of the boundary of the transition surface between these two 
Haldane phases with the ferromagnetic phase corresponds
to the exactly-soluble DKO model. We expect that this result will
suscitate further theoretical work on this model, concerning in particular
remarkable points like $\Gamma$ and C' on the DKO line 
(see Figs.~18 and 23).

Most  calculations were performed on the Cray T3E from CEA-Grenoble. 
Numerical data concerning the transition surfaces are available upon 
request\cite{michel}

\vfill\eject

\widetext

\begin{references}
%
\bibitem[*]{tigran} Permanent address: Yerevan Physics Institute, Alikhanian
br. 2, Yerevan Armenia.
%
\bibitem[\dag]{jack} Permanent address: Michigan State University, Physics
Department, East Lansing, 48824 MI USA.
%
\bibitem[\ddag]{michel} To whom correspondence should be addressed.
Electronic address: roger@drecam.saclay.cea.fr
%
\bibitem{dagotto}
E. Dagotto and T. M. Rice, Science {\bf 271}, 618 (1996).
%
\bibitem{bethe}
H. Bethe, Z. Phys. {\bf 71}, 205 (1931).
%
\bibitem{haldane}
F. D. M. Haldane, Phys. Rev. Lett. {\bf 50}, 1153 (1983).
%
\bibitem{white1}
S. R. White, Phys. Rev. B, {\bf 53}, 52 (1996).
%
\bibitem{wang}
Xiaoqun Wang, in {\it Density-Matrix Renormalization,
Lecture Notes in Physics}, edited by I. Peschel, X. Wang, M. Kaulke and
K. Hallberg, Springer (1999); Xiaoqun Wang, cond-mat/9803290 (unpublished).
%
\bibitem{white}
S. R. White, Phys. Rev. Lett. {\bf 69}, 2863 (1992); Phys. Rev. B {\bf 48}
10345 (1993).
%
\bibitem{aklt}
I. Affleck, T. Kennedy, E. H. Lieb, and H. Tasaki, Phys. Rev. Lett. {\bf 59},
799 (1987).
%
\bibitem{anderson}
P. W. Anderson, Mater. Res. Bull. {\bf 8}, 153 (1973).
%
\bibitem{rommelse}
K. Rommelse and M. den Nijs, Phys. Rev. Lett. {\bf 59}, 2578 (1987).
%
\bibitem{kim}
E. H. Kim, G. F\'ath, J. S\'olyom and D.J. Scalapino, cond-mat/9910023.
%
\bibitem{kolezhuk}
A. Kolezhuk, R. Roth and U. Schollw\"ock, Phys. Rev. Lett. {\bf 77}, 5142
(1996).
%
\bibitem{spin1}
S. R. White and D. A. Huse, Phys. Rev. B {\bf 48} 3844 (1993).
%
\bibitem{mg}
C. K. Majumdar and D. K. Ghosh, J. Math. Phys. {\bf 10}, 1388 (1969).
%
\bibitem{shastry}
B. S. Shastry and B. Sutherland, Phys. Rev. Lett. {\bf 47}, 964 (1981).
%
\bibitem{chitra}
R. Chitra, Swapan Pati, H. R. Krishnamurthy, Diptiman Sen and 
S. Ramasesha, Phys. Rev. B {\bf 52}, 6581 (1985).
%
\bibitem{neugebauer1}
S. Brehmer, H.-J. Mikeska and U. Neugebauer, J. Phys. Cond. Mat. {\bf 8},
7161 (1996).
%
\bibitem{emery}
V. J. Emery and C. Noguera, Phys. Rev. Lett. {\bf 60}, 631 (1988).
G. Castilla, S. Chakravarty and V. J. Emery, ibid. {\bf 75}, 1823 (1995).
%
\bibitem{eggert}
S. Eggert, Phys. Rev. B {\bf 54} R9612 (1999).
%
\bibitem{ovchinnikov}
D. V. Dmitriev, V. Ya Krivnov, and A. A. Ovchinnikov, Phys. Rev. B {\bf 56},
5985 (1997); cond-mat/9911394 (1999).
%
\bibitem{note} In Ref. 10, another view of ``universality'' is adopted,
in which two phases which differ topologically  are said to belong to 
different universality classes, even when they correspond through a simple
reflection symmetry.
%
\bibitem{km} A. K. Kohlezhuk and H.-J Mikeska, Int. J. Mod. Phys. B {\bf 12},
2325 (1998).
%
\bibitem{composite}
J. S\'olyom and J. Timonen, Phys. Rev. B {\bf 34}, 487 (1986); and references
therein.
%
\bibitem{sierra1} M.A. Martin-Delgado, R. Shankar, and G. Sierra, Phys. Rev.
Lett. {\bf 77}, 3443 (1996).
%
\bibitem{sierra2} M.A. Martin-Delgado, J. Dukelsky and G. Sierra,
Phys. Lett. A {\bf 250}, 430 (1998).


%
\end{references}
\end{document}